\documentclass[10pt]{article}
\usepackage{graphicx}
\usepackage{amsmath}
\usepackage{amssymb}
\usepackage{caption2}
\begin{document}
\bibliographystyle{prsty}
\begin{center}
{\large {\bf \sc{  The hadronic coupling constants of the lowest hidden-charm pentaquark  state  with the QCD sum rules in rigorous quark-hadron duality }}} \\[2mm]
Zhi-Gang Wang \footnote{E-mail: zgwang@aliyun.com.  }, Hui-Juan Wang, Qi Xin     \\
 Department of Physics, North China Electric Power University, Baoding 071003, P. R. China
\end{center}

\begin{abstract}
In this article, we illustrate how to calculate the hadronic coupling constants  of the pentaquark states with the QCD sum rules based on rigorous quark-hadron quality, then study the hadronic coupling constants of the lowest diquark-diquark-antiquark type hidden-charm  pentaquark state with the spin-parity $J^P={\frac{1}{2}}^-$  in details, and calculate the partial decay widths. The total  width $\Gamma(P_c)=14.32\pm3.31\,\rm{MeV}$ is compatible with the experimental value   $\Gamma_{P_c(4312)} = 9.8\pm2.7^{+ 3.7}_{- 4.5} \mbox{ MeV}$ from the LHCb collaboration, and  favors  assigning the $P_c(4312)$ to be the  $[ud][uc]\bar{c}$   pentaquark state with the $J^P={\frac{1}{2}}^-$. The  hadronic coupling constants have the relation $|G_{PD^-\Sigma_c^{++}}|= \sqrt{2}|G_{P\bar{D}^0\Sigma_c^+}|\gg |G_{P\bar{D}^0\Lambda_c^+}|$, and favor the hadronic dressing mechanism. The $P_c(4312)$ maybe have a diquark-diquark-antiquark type pentaquark core with the typical size of the $qqq$-type   baryon states, the strong couplings  to the meson-baryon pairs $\bar{D}\Sigma_c$  lead to some pentaquark molecule  components, and the $P_c(4312)$ maybe spend  a rather  large time as the $\bar{D}\Sigma_c$  molecular state.
\end{abstract}



\section{Introduction}

In 2015,  the  LHCb collaboration   observed  two pentaquark or pentaquark molecule candidates $P_c(4380)$ and $P_c(4450)$ in the $J/\psi p$ invariant  mass spectrum  in the $\Lambda_b^0\to J/\psi K^- p$ decays \cite{LHCb-4380}. In 2019, also in the $ J/\psi  p$ invariant mass spectrum, the LHCb collaboration  observed a  new narrow pentaquark or pentaquark molecule candidate $P_c(4312)$  and confirmed the old structure $P_c(4450)$, which consists  of two narrow overlapping peaks $P_c(4440)$ and $P_c(4457)$ \cite{LHCb-Pc4312}.
There have been several possible interpretations for the quark structures  of the  $P_c(4312)$, $P_c(4440)$ and $P_c(4457)$, such as  the pentaquark molecular states \cite{mole-penta-2015-1,mole-penta-2015-2,mole-penta-2015-3,mole-penta-2015-4,mole-penta-2015-5,mole-penta-2015-6,mole-penta-2015-7,mole-penta-2015-8,mole-penta-2015-9,
mole-penta-2015-10,WangPc4450-molecule,Pc4312-mole-penta-1,Pc4312-mole-penta-2,Pc4312-mole-penta-3,Pc4312-mole-penta-4,Pc4312-mole-penta-5,Pc4312-mole-penta-6,Pc4312-mole-penta-7,
Pc4312-mole-penta-8,Pc4312-mole-penta-9,Pc4312-mole-penta-10,Pc4312-mole-penta-11,Pc4312-mole-penta-12,Pc4312-mole-penta-13,Pc4312-mole-penta-14,Pc4312-mole-penta-15,Pc4312-Wang-mole-decay, Pc4312-LYL-mole-decay},
 compact diquark-diquark-antiquark type pentaquark states or diquark-triquark type pentaquark states \cite{di-di-anti-penta-2015-1,di-di-anti-penta-2015-2,di-di-anti-penta-2015-3,di-di-anti-penta-2015-4,di-di-anti-penta-2015-5,Wang1508-EPJC-1,
 Wang1508-EPJC-2,Wang1508-EPJC-3,Wang1508-EPJC-4,Wang-Huang-1508-EPJC,Pc4312-di-di-anti-1,Pc4312-di-di-anti-2,Pc4312-di-di-anti-3,WangZG-penta-2020-IJMPA,di-tri-penta-2015-1,di-tri-penta-2015-2,di-tri-penta-2015-3},  color-octet-color-octet type pentaquark states \cite{Octet-Lee},  hadrocharmonium pentaquark states \cite{Pc4312-hadrocharmonium}, etc.

In Refs.\cite{Wang1508-EPJC-1,Wang1508-EPJC-2,Wang1508-EPJC-3,Wang1508-EPJC-4,Wang-Huang-1508-EPJC}, we perform comprehensive investigations  of  the spin-parity $J^P={\frac{1}{2}}^\pm$, ${\frac{3}{2}}^\pm$ and ${\frac{5}{2}}^\pm$  diquark-diquark-antiquark type hidden-charm pentaquark  states
with the QCD sum rules by carrying out the operator product expansion   up to   the vacuum condensates of dimension $10$ in a consistent way and separating the contributions of the positive parity pentaquark states from that of the negative parity pentaquark states explicitly, and reproduce the experimental values of the masses of the $P_c(4380)$  and $P(4450)$ as the compact pentaquark states with the spin-parity $J^P={\frac{3}{2}}^-$ and ${\frac{5}{2}}^+$, respectively. Furthermore, we obtain the lowest  masses $4.29 \pm 0.13\,\rm{GeV}$ and $4.30 \pm 0.13\,\rm{GeV}$  for the scalar-diquark--scalar-diquark-antiquark type and scalar-diquark-axialvector-diquark-antiquark type hidden-charm pentaquark  states with the spin-parity  $J^P={\frac{1}{2}}^-$, respectively \cite{Wang-Huang-1508-EPJC}, which are all consistent with the mass of the $P_c(4312)$ observed later by the LHCb collaboration \cite{LHCb-Pc4312}. Then we update the old analysis by   taking into account
   the vacuum condensates up to dimension $13$ in a consistent way \cite{WangZG-penta-2020-IJMPA}, and obtain more flatter Borel platforms and better predictions of the masses and pole residues. The new analysis indicates that the lowest scalar-diquark--scalar-diquark-antiquark type and axialvector-diquark-axialvector-diquark-antiquark type compact hidden-charm pentaquark  states with the spin-parity $J^P={\frac{1}{2}}^-$ have the masses $4.31 \pm 0.11\,\rm{GeV}$ and $4.34 \pm 0.14\,\rm{GeV}$, respectively, which are all consistent with the mass of the $P_c(4312)$. While the scalar-diquark-axialvector-diquark-antiquark type hidden-charm pentaquark  states with the spin-parity  $J^P={\frac{1}{2}}^-$ has a mass $4.45 \pm 0.11\,\rm{GeV}$ rather than $4.30 \pm 0.13\,\rm{GeV}$ \cite{WangZG-penta-2020-IJMPA}.

On the other hand, in Ref.\cite{WangPc4450-molecule}, we perform detailed investigations  of  the $\bar{D}\Sigma_c$, $\bar{D}\Sigma_c^*$, $\bar{D}^{*}\Sigma_c$  and   $\bar{D}^{*}\Sigma_c^*$  pentaquark molecular states
with the QCD sum rules by carrying out the operator product expansion   up to   the vacuum condensates of dimension $13$ in a consistent way.
 The theoretical predictions of the molecule masses favor assigning the $P_c(4312)$ to be the $\bar{D}\Sigma_c$ pentaquark molecular state with the spin-parity $J^P={\frac{1}{2}}^-$, assigning the $P_c(4380)$ to be the $\bar{D}\Sigma_c^*$ pentaquark molecular state with the spin-parity $J^P={\frac{3}{2}}^-$, and assigning the $P_c(4440/4457)$ to be the $\bar{D}^{*}\Sigma_c$ pentaquark molecular state with the spin-parity $J^P={\frac{3}{2}}^-$ or the $\bar{D}^{*}\Sigma_c^*$ pentaquark molecular state with the spin-parity $J^P={\frac{5}{2}}^-$, respectively. In the works of other theoretical groups, the  $P_c(4312)$, $P_c(4380)$, $P_c(4440)$ and $P_c(4457)$ are taken as the pentaquark molecular states, and their masses are studied with the QCD sum rules by carrying out the operator product expansion up to the vacuum condensates of the dimension $8$ \cite{mole-penta-2015-2,Pc4312-mole-penta-2,Pc4312-mole-penta-10} or $6$ \cite{mole-penta-2015-9}. The works on the decay widths of the $P_c(4312)$, $P_c(4380)$, $P_c(4440)$ and $P_c(4457)$ are few, in Ref.\cite{Pc4312-Wang-mole-decay} and Ref.\cite{Pc4312-LYL-mole-decay}, the $P_c(4312)$ is assigned to be the pentaquark molecular state, its  two-body strong decays  are studied with the QCD sum rules by carrying out the operator product expansion up to the vacuum condensates of dimension $10$ and $8$,  respectively; while in Ref.\cite{mole-penta-2015-10}, the $P_c(4380)$ is assigned to be the pentaquark molecular state, its  two-body strong decays  are studied with the QCD sum rules by carrying out the operator product expansion up to the vacuum condensates of dimension $6$.

It is odd that the experimental values of the masses of the  $P_c(4312)$, $P_c(4440)$ and $P_c(4457)$ can be reproduced both in the scenarios of the pentaquark  states and pentaquark molecular states with the QCD sum rules. A hadron has definite quantum numbers and several Fock states, any current with the same quantum numbers and quark structures as a Fock state in a hadron couples potentially to this  hadron. In this respect, we can construct several currents to interpolate a hadron, or construct a current to interpolate several hadrons. However, we should bear in mind that a hadron has one or two main Fock states, we call a hadron as a pentaquark (molecular) state if its main Fock component is of the diquark-diquark-antiquark type (color-singlet-color-singlet type),   and try to choose the pertinent current to interpolate it. In the present case, the diquark-diquark-antiquark type local pentaquark current   with definite  quantum numbers couples potentially  to a definite compact pentaquark state, though this  local current  can be re-arranged into a special superposition of  a series of color-singlet-color-singlet type  currents, which couple potentially  to the pentaquark molecular states or meson-baryon two-hadron scattering  states  with the same quantum numbers \cite{WangZG-penta-2020-IJMPA}.
The diquark-diquark-antiquark type pentaquark states can be taken as a special superposition of  a series of color-singlet-color-singlet molecular states and embody the net effects, and vise versa.

We can borrow some ideas from the nature of the light flavor scalar mesons, which provide a subject of an intense and continual  controversy in establishing  the meson spectrum, the most  elusive things are the quark configurations of
 the $f_0(980)$ and $a_0(980)$, which have  almost the degenerate masses.  In the scenario of the hadronic dressing mechanism,  the scalar  mesons $f_0(980)$ and $a_0(980)$ have  small or large  $q\bar{q}$ cores of the typical  $q\bar{q}$ meson size, or large $[qq]_{\bar{3}}[\bar{q}\bar{q}]_3$ cores in the relative S-wave with some $q\bar{q}$ components  in the relative P-wave,
 the bare $q\bar{q}$ or $[qq]_{\bar{3}}[\bar{q}\bar{q}]_3$ cores are dressed by the hadronic interactions with the pseudoscalar mesons,
   the strong couplings to the   hadronic
channels or nearby thresholds enrich the pure $q\bar{q}$ or $[qq]_{\bar{3}}[\bar{q}\bar{q}]_3$ states with other components and spend part or most part of their lifetime as the virtual
$K^+K^-$ or $\bar{K}^0K^0$ states \cite{Hadron-dress-1,Hadron-dress-2,Close2002,ReviewAmsler2,PColangelo-2003-1,PColangelo-2003-2}. The QCD sum rules indicate that
 the nonet scalar mesons  below $1\,\rm{ GeV}$ are the two-quark-tetraquark  mixing  states with large or small  two-quark components \cite{WangZG-lightScalar-EPJC,tetra-qq}.  Without introducing mixing effects in one way or the other, it is difficult to reproduce the experimental values of the masses of the nonet scalar mesons  below $1\,\rm{ GeV}$ \cite{Lee2006,LeeHJ-2019}, and account for the decays.
 In summary, the QCD sum rules favor the hadronic dressing mechanism  \cite{PColangelo-2003-1,PColangelo-2003-2,WangZG-lightScalar-EPJC,tetra-qq}.

The hadronic dressing mechanism also works in interpreting the exotic $X$, $Y$ and $Z$ states. In Ref.\cite{WangZG-solid-4660},  we  choose  the $[sc]_P[\bar{s}\bar{c}]_A-[sc]_A[\bar{s}\bar{c}]_P$  type tetraquark current to study the hadronic coupling constants in the  strong decays of the  $Y(4660)$ with the QCD sum rules based on rigorous quark-hadron quality. The numerical values indicate that  the  hadronic coupling constants $ |G_{Y\psi^\prime f_0}|\gg |G_{Y J/\psi f_0}|$, which is consistent with the fact that  the $Y(4660)$ is observed in the $\psi^\prime\pi^+\pi^-$ invariant mass distribution, and favors the $\psi^{\prime}f_0(980)$ molecule assignment considering the decay chains $Y(4600)\to \psi^{\prime}f_0(980) \to \psi^{\prime}\pi^+\pi^-$ \cite{Wang-CTP-4660-1,Wang-CTP-4660-2}. Similar  mechanism maybe exist for the pentaquark states and pentaquark molecular states, i.e. the pentaquark states  maybe have a diquark-diquark-antiquark type pentaquark core with the typical size of the $qqq$-type baryon states, the strong couplings to the meson-baryon pairs  lead to some pentaquark molecule Fock components, and the valance quarks are rearranged into the color-singlet-color-singlet periphery structures, and  spend a rather large time as the molecular states.

In the article, we study the hadronic coupling constants of the lowest scalar-diquark-scalar-diquark-antiquark type hidden-charm pentaquark state with the QCD sum rules base on the  rigorous  quark-hadron duality, and study its two-body strong decays and estimate the magnitude of the total  decay width, and examine the hadronic dressing mechanism for the compact pentaquark states, and try to compromise the scenarios of the pentaquark  states and pentaquark molecular states.

 The article is arranged as follows:  in Sect.2, we illustrate how to calculate the hadronic coupling constants of the hidden-charm pentaquark states with the QCD sum rules based on the rigorous quark-hadron quality; in Sect.3, we derive the QCD sum rules for the  hadronic  coupling constants of the lowest hidden-charm pentaquark state with the spin-parity $J^P={\frac{1}{2}}^-$;   in Sect.4, we present the numerical results and discussions; and Sect.5 is reserved for our
conclusion.

\section{The hadronic coupling constants  of the hidden-charm pentaquark states  }

In this section, we illustrate how to calculate the hadronic coupling constants of  the hidden-charm pentaquark states with the QCD sum rules.
Firstly, let us write down the  three-point correlation functions $\Pi(p,q)$,
\begin{eqnarray}
\Pi(p,q)&=&i^2\int d^4xd^4y e^{ip \cdot x}e^{iq \cdot y}\langle 0|T\left\{J_{M}(x)J_{B}(y)\bar{J}_{P}(0)\right\}|0\rangle\, ,
\end{eqnarray}
where $\bar{J}_P(0)=J_P^\dagger(0)\gamma^0$, the current $J_P(0)$ interpolates the hidden-charm pentaquark state $P_c$, the $J_M(x)$ and $J_B(y)$ interpolate the  traditional  meson $M$ and baryon $B$,  respectively,
\begin{eqnarray}
\langle0|J_{P}(0)|P_c(p^\prime)\rangle&=&\lambda_{P}U(p^\prime,s) \,\, , \nonumber \\
\langle0|J_{M}(0)|M(p)\rangle&=&\lambda_{M} \,\, , \nonumber \\
\langle0|J_{B}(0)|B(q)\rangle&=&\lambda_{B}U(q,s) \,\, ,
\end{eqnarray}
the $\lambda_P$, $\lambda_M$  and $\lambda_{B}$ are the pole residues or   decay constants, the $U(p^\prime,s)$ and $U(q,s)$ are the Dirac spinors.

At the hadron  side,  we insert  a complete set of intermediate hadronic states with
the same quantum numbers as the current operators $\bar{J}_{P}(0)$, $J_{M}(x)$, $J_{B}(y)$ into the three-point
correlation functions  $\Pi(p,q)$ and  isolate the ground state
contributions of the pentaquark state $P_c$, traditional  meson $M$ and baryon $B$ to obtain the  hadronic representation  \cite{SVZ79,PRT85},
\begin{eqnarray}
\Pi(p,q)&=&-i\lambda_{P}\lambda_{M}\lambda_{B} \frac{ \left(\!\not\!{q}+m_B \right)G_{PMB} \Gamma\left(\!\not\!{p}^\prime+m_P \right) }{(m_{M}^2-p^2)(m_{B}^2-q^2)(m_{P}^2-p^{\prime2})}+\cdots\, ,
\end{eqnarray}
where $p^\prime=p+q$,  the $G_{PMB}$  are the hadronic coupling constants defined by
\begin{eqnarray}
\langle M(p)B(q)|P_c(p^{\prime})\rangle&=& G_{PMB}\overline{U}(q)\Gamma U(p^\prime)   \, ,
\end{eqnarray}
the $\Gamma$  are some Dirac
$\gamma$-matrixes.

In the QCD sum rules, irrespective of the two-point or three-point QCD sum rules, we take the quark-hadron duality to match the hadron representation with the  QCD representation of the correlation functions,
\begin{eqnarray}
\Pi_{H}(p,q)&=&\Pi_{QCD}(p,q)\, ,
\end{eqnarray}
 where we add the subscripts $H$ and $QCD$ to denote the  hadron side and   QCD side, respectively.
We expect the  equality,
 \begin{eqnarray}\label{Tr-Gamma}
\frac{1}{4}{\rm Tr}\left[\Pi_{H}(p,q)\Gamma^\prime\right]&=&\frac{1}{4}{\rm Tr}\left[\Pi_{QCD}(p,q)\Gamma^\prime\right]\, ,
\end{eqnarray}
survives after multiplying both sides by $\Gamma^\prime$ and accomplishing  the trace in the Dirac spinor space, where  the $\Gamma^\prime$  are some Dirac
$\gamma$-matrixes.
If we choose $\Gamma=1$ and $\Gamma^\prime=\sigma_{\mu\nu}$, then we obtain
  \begin{eqnarray}
\frac{1}{4}{\rm Tr}\left[\Pi_{H}(p,q)\sigma_{\mu\nu}\right]&=&\Pi_{H}(p^{\prime2},p^2,q^2) \left(p_{\mu}q_{\nu}-p_{\nu}q_{\mu} \right)\, , \nonumber\\
\frac{1}{4}{\rm Tr}\left[\Pi_{QCD}(p,q)\sigma_{\mu\nu}\right]&=&\Pi_{QCD}(p^{\prime2},p^2,q^2) \left(p_{\mu}q_{\nu}-p_{\nu}q_{\mu} \right)\, ,
\end{eqnarray}
where the  $\Pi_{H}(p^{\prime2},p^2,q^2)$ and $\Pi_{QCD}(p^{\prime2},p^2,q^2)$ are the relevant components of the correlation functions $\Pi(p,q)$ we want to study at the hadron side and QCD side, respectively. Let us write down the components  $\Pi_{H}(p^{\prime2},p^2,q^2)$ explicitly,
\begin{eqnarray}\label{Compare-2005-H}
\Pi_{H}(p^{\prime2},p^2,q^2)&=&  \frac{ \lambda_{P}\lambda_{M}\lambda_{B}G_{PMB} }{(m_{M}^2-p^2)(m_{B}^2-q^2)(m_{P}^2-p^{\prime2})}\nonumber\\
&&+ \frac{1}{(m_{M}^2-p^2)(m_{P}^2-p^{\prime2})} \int_{s^0_B}^\infty dt\frac{\rho_{PB^\prime}(p^{\prime 2},p^2,t)}{t-q^2}\nonumber\\
&& + \frac{1}{(m_{B}^2-q^2)(m_{P}^2-p^{\prime2})} \int_{s^0_{M}}^\infty dt\frac{\rho_{PM^\prime}(p^{\prime 2},t,q^2)}{t-p^2}  \nonumber\\
&& + \frac{1}{(m_{M}^2-p^{2})(m_{B}^2-q^2)} \int_{s^0_{P}}^\infty dt\frac{\rho_{P^{\prime}M}(t,p^2,q^2)+\rho_{P^{\prime}B}(t,p^2,q^2)}{t-p^{\prime2}}+\cdots  \, ,
\end{eqnarray}
where we introduce the four  formal  functions $\rho_{PB^\prime}(p^{\prime 2},p^2,t)$, $ \rho_{PM^\prime}(p^{\prime 2},t,q^2)$,
$ \rho_{P^{\prime}M}(t^\prime,p^2,q^2)$ and $\rho_{P^{\prime}B}(t^\prime,p^2,q^2)$ to
   parameterize  the complex  couplings or transitions between the ground states and the higher resonances  or the continuum states.
 In Ref.\cite{mole-penta-2015-10}, the $P_c(4380)$ is assigned to be  a pentaquark molecular state, its two-body strong decays  are studied with the QCD sum rules, where the second term, the third term and  the fourth term in Eq.\eqref{Compare-2005-H} are all neglected.
   In Refs.\cite{Pc4312-Wang-mole-decay,Pc4312-LYL-mole-decay}, the $P_c(4312)$ is assigned to be  a pentaquark molecular state, its two-body strong decays  are studied with the QCD sum rules, all the terms in Eq.\eqref{Compare-2005-H} are taken into account, just as what was  suggested in Ref.\cite{Pc4312-Wang-mole-decay}.
   In Refs.\cite{Nilesen-P-2005,WangZG-P-2005}, the $\Theta^+(1540)$ is assigned to be a pentaquark state, its two-body strong decays   are studied with the QCD sum rules, where the second term, the third term or  the fourth term are neglected in one way or the other. We should take into account all the four terms in Eq.\eqref{Compare-2005-H} so as to  describe the transitions between the ground states and the first radial excites  in order  to make the calculation robust.

We rewrite the correlation functions  $\Pi_H(p^{\prime 2},p^2,q^2)$ at the hadron  side as
\begin{eqnarray}
\Pi_{H}(p^{\prime 2},p^2,q^2)&=&\int_{(m_{M}+m_{B})^2}^{s_{P}^0}ds^\prime \int_{\Delta_s^2}^{s^0_{M}}ds \int_{\Delta_u^2}^{s^0_{B}}du  \frac{\rho_H(s^\prime,s,u)}{(s^\prime-p^{\prime2})(s-p^2)(u-q^2)}\nonumber\\
&&+\int_{s^0_P}^{\infty}ds^\prime \int_{\Delta_s^2}^{s^0_{M}}ds \int_{\Delta_u^2}^{s^0_{B}}du  \frac{\rho_H(s^\prime,s,u)}{(s^\prime-p^{\prime2})(s-p^2)(u-q^2)}+\cdots\, ,
\end{eqnarray}
 through triple  dispersion relation, where the $\rho_H(s^\prime,s,u)$   are the hadronic spectral densities,
\begin{eqnarray}\label{Hadron-Spectral-Denti}
\rho_H(s^\prime,s,u)&=&{\lim_{\epsilon_3\to 0}}\,\,{\lim_{\epsilon_2\to 0}} \,\,{\lim_{\epsilon_1\to 0}}\,\,\frac{ {\rm Im}_{s^\prime}\, {\rm Im}_{s}\,{\rm Im}_{u}\,\Pi_H(s^\prime+i\epsilon_3,s+i\epsilon_2,u+i\epsilon_1) }{\pi^3} \, ,
\end{eqnarray}
where the $\Delta_s^2$ and $\Delta_u^2$ are the thresholds in the $s$ and $u$ channels, respectively,  the  $s_{P}^0$, $s_{M}^0$, $s_{B}^0$ are the continuum threshold parameters.

Now we carry out the operator product expansion at the QCD side in the deep Euclidean  region $P^2=-p^2\gg \Lambda^2_{QCD}$ and $Q^2=-q^2\gg \Lambda^2_{QCD}$.
However,   we cannot write the correlation functions  $\Pi_{QCD}(p^{\prime 2},p^2,q^2)$ in the form,
\begin{eqnarray}
\Pi_{QCD}(p^{\prime 2},p^2,q^2)&=&\int_{(m_{M}+m_{B})^2}^{s_{P}^0}ds^\prime \int_{\Delta_s^2}^{s^0_{M}}ds \int_{\Delta_u^2}^{s^0_{B}}du  \frac{\rho_{QCD}(s^\prime,s,u)}{(s^\prime-p^{\prime2})(s-p^2)(u-q^2)}\nonumber\\
&&+\int_{s^0_P}^{\infty}ds^\prime \int_{\Delta_s^2}^{s^0_{M}}ds \int_{\Delta_u^2}^{s^0_{B}}du  \frac{\rho_{QCD}(s^\prime,s,u)}{(s^\prime-p^{\prime2})(s-p^2)(u-q^2)}+\cdots\, ,
\end{eqnarray}
through triple dispersion relation analogously, because the QCD spectral densities $\rho_{QCD}(s^\prime,s,u)$ cannot exist,
\begin{eqnarray}
\rho_{QCD}(s^\prime,s,u)&=&{\lim_{\epsilon_3\to 0}}\,\,{\lim_{\epsilon_2\to 0}} \,\,{\lim_{\epsilon_1\to 0}}\,\,\frac{ {\rm Im}_{s^\prime}\, {\rm Im}_{s}\,{\rm Im}_{u}\,\Pi_{QCD}(s^\prime+i\epsilon_3,s+i\epsilon_2,u+i\epsilon_1) }{\pi^3} \nonumber\\
&=&0\, ,
\end{eqnarray}
and we have to write the correlation functions  $\Pi_{QCD}(p^{\prime 2},p^2,q^2)$  in the form,
\begin{eqnarray}
\Pi_{QCD}(p^{\prime 2},p^2,q^2)&=&  \int_{\Delta_s^2}^{s^0_{M}}ds \int_{\Delta_u^2}^{s^0_{B}}du  \frac{\rho_{QCD}(p^{\prime2},s,u)}{(s-p^2)(u-q^2)}+\cdots\, ,
\end{eqnarray}
through double dispersion relation, where the $\rho_{QCD}(p^{\prime 2},s,u)$   are the QCD spectral densities,
\begin{eqnarray}
\rho_{QCD}(p^{\prime 2},s,u)&=& {\lim_{\epsilon_2\to 0}} \,\,{\lim_{\epsilon_1\to 0}}\,\,\frac{  {\rm Im}_{s}\,{\rm Im}_{u}\,\Pi_{QCD}(p^{\prime 2},s+i\epsilon_2,u+i\epsilon_1) }{\pi^2} \, .
\end{eqnarray}
Henceforth  we will write  the QCD spectral densities  $\rho_{QCD}(p^{\prime 2},s,u)$ in the form  $\rho_{QCD}(s,u)$ for simplicity.

As the duality below the three continuum threshold parameters $s_P^0$, $s_M^0$ and $s_B^0$  cannot exist simultaneously,
\begin{eqnarray}
&&\int_{(m_{M}+m_{B})^2}^{s_{P}^0}ds^\prime \int_{\Delta_s^2}^{s^0_{M}}ds \int_{\Delta_u^2}^{s^0_{B}}du  \frac{\rho_H(s^\prime,s,u)}{(s^\prime-p^{\prime2})(s-p^2)(u-q^2)}\nonumber\\
&\neq&\int_{(m_{M}+m_{B})^2}^{s_{P}^0}ds^\prime \int_{\Delta_s^2}^{s^0_{M}}ds \int_{\Delta_u^2}^{s^0_{B}}du  \frac{\rho_{QCD}(s^\prime,s,u)}{(s^\prime-p^{\prime2})(s-p^2)(u-q^2)}\, ,
\end{eqnarray}
we  carry out the formal integral over $ds^\prime$  firstly, then we match the hadron side  with the QCD side of the correlation functions $\Pi(p^{\prime2},p^2,q^2)$ below the
 two continuum threshold parameters $s_M^0$ and $s_B^0$ simultaneously
   to obtain the rigorous  duality \cite{Pc4312-Wang-mole-decay,WangZG-solid-4660,WangZhang-Solid-1,WangZhang-Solid-2,WangZhang-Solid-3,WangZhang-Solid-4},
\begin{eqnarray}\label{duality}
\int_{\Delta_s^2}^{s_M^0} ds \int_{\Delta_u^2}^{s_B^0} du \frac{1}{(s-p^2)(u-q^2)}\left[ \int_{\Delta^2}^{\infty} ds^\prime \frac{\rho_{H}(s^{\prime},s,u)}{s^\prime-p^{\prime2}}\right]&=&\int_{\Delta_s^2}^{s_M^0} ds \int_{\Delta_u^2}^{s_B^0} du \frac{\rho_{QCD}(s,u)}{(s-p^2)(u-q^2)}\, , \nonumber\\
\end{eqnarray}
 where  $\Delta^2=(m_{M}+m_{B})^2$. We carry out the integral $\int_{\Delta^2}^{\infty} ds^\prime \frac{\rho_{H}(s^{\prime},s,u)}{s^\prime-p^{\prime2}}$ according to
 the hadronic spectral densities in Eq.\eqref{Hadron-Spectral-Denti} to make the calculations rigorous or robust, rather than just selecting or modeling the hadron  representations  by hand as in Refs.\cite{mole-penta-2015-10,Nilesen-P-2005,WangZG-P-2005}. Now let us  write down  the quark-hadron duality explicitly,
 \begin{eqnarray}\label{quark-hadron-duality}
  \int_{\Delta_s^2}^{s^0_{M}}ds \int_{\Delta_u^2}^{s^0_{B}}du  \frac{\rho_{QCD}(s,u)}{(s-p^2)(u-q^2)}&=& \int_{\Delta_s^2}^{s^0_{M}}ds \int_{\Delta_u^2}^{s^0_{B}}du   \int_{\Delta^2}^{\infty}ds^\prime \frac{\rho_H(s^\prime,s,u)}{(s^\prime-p^{\prime2})(s-p^2)(u-q^2)} \nonumber\\
  &=&\frac{\lambda_{P}\lambda_{M}\lambda_{B}G_{PMB} }{(m_{P}^2-p^{\prime2})(m_{M}^2-p^2)(m_{B}^2-q^2)} +\frac{C_{P^{\prime}M}+C_{P^{\prime}B}}{(m_{M}^2-p^{2})(m_{B}^2-q^2)} \, , \nonumber\\
\end{eqnarray}
 where we introduce the parameters  $C_{P^\prime M}$ and $C_{P^\prime B}$   to parameterize the net effects by neglecting the dependence on the variables $t$, $p^{\prime2}$, $p^2$ and $q^2$,
\begin{eqnarray}\label{subtract-constants}
C_{P^\prime M}&=&\int_{s^0_{P}}^\infty dt\frac{ \rho_{P^\prime M}(t,p^2,q^2)}{t-p^{\prime2}}\, ,\nonumber\\
C_{P^\prime B}&=&\int_{s^0_{P}}^\infty dt\frac{ \rho_{P^\prime B}(t,p^2,q^2)}{t-p^{\prime2}}\, .
\end{eqnarray}
 From Eq.\eqref{quark-hadron-duality}, we can see that the duality below the continuum threshold parameters $s_M^0$ and $s_B^0$ is rigorous.

In Eqs.\eqref{duality}-\eqref{quark-hadron-duality}, the continuum threshold parameters $s^0_{M}$ and $s^0_{B}$   appear both in the hadron side and QCD side of the correlation functions. As the spectroscopy of the traditional mesons and baryons are known much better than that of the pentaquark states, even the pentaquark states have not been established yet, we can consult the experimental data from the Particle Data Group and the theoretical  predictions from the two-point QCD sum rules to choose suitable continuum threshold parameters $s^0_{M}$ and $s^0_{B}$,
 which should be large enough to include the  contributions of the ground states fully, but small enough to exclude the contaminations of the
  higher excited states and continuum states.

In the two-point QCD sum rules for the conventional baryons $B$ and  mesons $M$,
\begin{eqnarray}\label{residue-QCDSR}
\lambda_{i}^2\exp\left(-\tau m_i^2 \right)&=& \int_{\Delta_i^2}^{s_i^0} ds \,\rho_{QCD}(s)\exp\left( -\tau s\right)\, ,
\end{eqnarray}
\begin{eqnarray}\label{mass-QCDSR}
m_i^2&=& \frac{-\frac{d}{d\tau}\int_{\Delta_i^2}^{s_i^0} ds \,\rho_{QCD}(s)\exp\left( -\tau s\right)}{\int_{\Delta_i^2}^{s_i^0} ds \,\rho_{QCD}(s)\exp\left( -\tau s\right)}\, ,
\end{eqnarray}
where $i=B$, $M$, $\tau=\frac{1}{T^2}$, the $T^2$ is the Borel parameter.  The predicted masses $m_{B/M}$ and pole residues $\lambda_{B/M}$ vary with the continuum threshold  parameters $s^0_{B/M}$. In calculations, we observe that the uncertainties  of the continuum threshold parameters,  $s^0_{i} \to s^0_{i}+\delta s^0_{i}$,  can lead to uncertainties of the masses and pole residues, $m_{i}\to m_{i}+\delta m_{i}$ and
$\lambda_{i}\to \lambda_{i}+\delta\lambda_{i}$,  with the relation $\frac{\delta\lambda_{i}}{\lambda_{i}}\gg \frac{\delta m_{i}}{m_{i}}$. In Eqs.\eqref{duality}-\eqref{quark-hadron-duality}, also in other three-point QCD sum rules for the hadronic coupling constants,  we usually take the physical masses $m_{P/B/M}$ as input parameters and neglect  the small  uncertainties. Furthermore, we can factorize out the pole residues $\lambda_{B/M}$ from the unknown  functions $C_{P^\prime M}=\lambda_{M}\lambda_{B}\widetilde{C}_{P^\prime M}$ and $C_{P^\prime B}=\lambda_{M}\lambda_{B}\widetilde{C}_{P^\prime B}$,
\begin{eqnarray}\label{quark-hadron-duality-rewrite}
  \int_{\Delta_s^2}^{s^0_{M}}ds \int_{\Delta_u^2}^{s^0_{B}}du  \frac{\rho_{QCD}(s,u)}{(s-p^2)(u-q^2)}&=&\lambda_{M}\lambda_{B}\left[ \frac{\lambda_{P}G_{PMB} }{(m_{P}^2-p^{\prime2})(m_{M}^2-p^2)(m_{B}^2-q^2)}\right.\nonumber\\
 &&  \left.+ \frac{\widetilde{C}_{P^{\prime}M}+\widetilde{C}_{P^{\prime}B}}{(m_{M}^2-p^{2})(m_{B}^2-q^2)}\right] \, ,
\end{eqnarray}
the uncertainties originate from the continuum threshold parameters $s^0_{M/B}$ can be absorbed into the pole residues approximately considering of the integrals $ds$ and $du$ in both sides of  Eqs.\eqref{duality}-\eqref{quark-hadron-duality}, the net effects due to the $\delta s^0_{B/M}$ are tiny.
We  choose the ideal values of the continuum threshold parameters $s^0_{M/B}$, which happen to reproduce the experimental values of the masses $m_{M/B}$ approximately, and neglect the uncertainties  for simplicity.

In fact, the QCD spectral densities $\rho_{QCD}(s,u)$ cannot be factorized out as $\rho_{QCD}(s)\,\rho_{QCD}(u)$, the continuum threshold parameters $s_0$ and $u_0$ are not necessary to  be the ones obtained from the two-point QCD sum rules. We can take the values obtained from  the two-point QCD sum rules as a guide, and vary the  $s_0$ and $u_0$ to search for the best parameters in the three-point QCD sum rules via trial and error. In calculations, we observe that the  deviations  $\delta s_0$ and $\delta u_0$ can be also compensated by variations of the parameters $\widetilde{C}_{P^{\prime}M}$ and $\widetilde{C}_{P^{\prime}B}$, if we fix the values of the pole residues $\lambda_{M}$ and $\lambda_{B}$. The  continuum threshold parameters $s_0$ and $u_0$  from the two-point QCD sum rules work well in the three-point QCD sum rules.

In numerical calculations,  we  can take the unknown  functions $C_{P^\prime M}$ and $C_{P^\prime B}$  as free parameters, and choose the suitable values  to account for
  the contaminations of the higher resonances and continuum states in the $s^\prime$ channel  to obtain the stable QCD sum rules. The parameters $C_{P^\prime M}$ and $C_{P^\prime B}$  are not necessary to be constants, they maybe depend on the Borel parameters, as
there are  complex interactions  or transitions between the ground states and the higher resonances  or  continuum states, after the double Borel transform, there maybe appear some net Borel parameter dependence.

If the  $M$  is a charmonium or bottomnium state and the $B$ is a light flavor baryon state, we set  $p^{\prime2}=p^2$  and perform the double Borel transform
in regard  to the variables $P^2=-p^2$ and $Q^2=-q^2$ respectively  to obtain the  QCD sum rules,
\begin{eqnarray}\label{QCDSR-4P2}
&& \frac{\lambda_{P}\lambda_{M}\lambda_{B}G_{PMB}}{m_{P}^2-m_{M}^2} \left[ \exp\left(-\frac{m_{M}^2}{T_1^2} \right)-\exp\left(-\frac{m_{P}^2}{T_1^2} \right)\right]\exp\left(-\frac{m_{B}^2}{T_2^2} \right) +\nonumber\\
&&\left(C_{P^{\prime}M}+C_{P^{\prime}B}\right) \exp\left(-\frac{m_{M}^2}{T_1^2} -\frac{m_{B}^2}{T_2^2} \right)=\int_{\Delta_s^2}^{s_M^0} ds \int_{\Delta_u^2}^{s_B^0} du\, \rho_{QCD}(s,u)\exp\left(-\frac{s}{T_1^2} -\frac{u}{T_2^2} \right)\, ,\nonumber\\
\end{eqnarray}
where the $T_1^2$ and $T_2^2$ are the Borel parameters. On the other hand, if the  $M$   is a heavy  meson and the $B$ is a heavy baryon state,  we set  $p^{\prime2}=4q^2$  and perform the double Borel transform  in regard  to the variables $P^2=-p^2$ and $Q^2=-q^2$ respectively  to obtain the  QCD sum rules,
\begin{eqnarray}\label{QCDSR-4Q2}
&& \frac{\lambda_{P}\lambda_{M}\lambda_{B}G_{PMB}}{4\left(\widetilde{m}_{P}^2-m_{B}^2\right)} \left[ \exp\left(-\frac{m_{B}^2}{T_2^2} \right)-\exp\left(-\frac{\widetilde{m}_{P}^2}{T_2^2} \right)\right]\exp\left(-\frac{m_{M}^2}{T_1^2} \right) +\nonumber\\
&&\left(C_{P^{\prime}M}+C_{P^{\prime}B}\right) \exp\left(-\frac{m_{B}^2}{T_2^2} -\frac{m_{M}^2}{T_1^2} \right)=\int_{\Delta_s^2}^{s_M^0} ds \int_{\Delta_u^2}^{s_B^0} du\, \rho_{QCD}(s,u)\exp\left(-\frac{s}{T_1^2} -\frac{u}{T_2^2} \right)\, , \nonumber\\
\end{eqnarray}
where $\widetilde{m}_P^2=\frac{m_P^2}{4}$.

In the QCD sum rules in Eqs.\eqref{QCDSR-4P2}-\eqref{QCDSR-4Q2}, the Borel parameters $T_1^2$ and $T_2^2$ are independent parameters,  the intervals of dimensions
of the vacuum condensates are small. In calculations, we can set $T_1^2=T_2^2=T^2$ to obtain  much larger intervals of dimensions of the vacuum condensates, therefore
  much stable QCD sum rules and much better accuracy of the predictions.

\section{QCD sum rules for  the hadronic coupling  constants of the lowest hidden-charm   pentaquark state with $J^P={\frac{1}{2}}^-$}

In the following, we write down  the three-point correlation functions $\Pi(p,q)$ and $\Pi_{\mu}(p,q)$  in the QCD sum rules,
\begin{eqnarray}
\Pi(p,q)&=&i^2\int d^4xd^4y e^{ip \cdot x}e^{iq \cdot y} \langle0|T\left\{J_M(x)J_B(y)\bar{J}_P(0)\right\}|0\rangle \, , \\
\Pi_\mu(p,q)&=&i^2\int d^4xd^4y e^{ip \cdot x}e^{iq \cdot y} \langle0|T\left\{J_\mu(x)J_N(y)\bar{J}_P(0)\right\}|0\rangle \, ,
\end{eqnarray}
where $J_M(x)= J_{\eta_c}(x)$, $J_{\bar{D}^0}(x)$, $J_{D^-}(x)$, $J_B(y)= J_{\Lambda_c^+}(y)$, $J_{\Sigma_c^{++}}(y)$, $J_{\Sigma_c^+}(y)$, $J_N(y)$,
\begin{eqnarray}
 J_{\eta_c}(x)&=& \bar{c}(x)i\gamma_5 c(x)\, ,\nonumber\\
 J_{\bar{D}^0}(x)&=& \bar{c}(x)i\gamma_5 u(x)\, ,\nonumber\\
 J_{D^-}(x)&=& \bar{c}(x)i\gamma_5 d(x)\, ,\nonumber\\
 J_\mu(x)&=& \bar{c}(x) \gamma_\mu c(x)\, ,
 \end{eqnarray}
 \begin{eqnarray}
  J_{\Lambda_c^+}(y)&=&  \varepsilon^{ijk}  u^T_i(y) C\gamma_5 d_j(y)\,  c_{k}(y) \, ,\nonumber \\
  J_{\Sigma_c^{++}}(y)&=&  \varepsilon^{ijk}  u^T_i(y) C\gamma_\alpha u_j(y)\, \gamma^\alpha\gamma_5 c_{k}(y) \, ,\nonumber \\
  J_{\Sigma_c^+}(y)&=&  \varepsilon^{ijk}  u^T_i(y) C\gamma_\alpha d_j(y)\, \gamma^\alpha\gamma_5 c_{k}(y) \, ,\nonumber \\
 J_N(y)&=&  \varepsilon^{ijk}  u^T_i(y) C\gamma_\alpha u_j(y)\, \gamma^\alpha\gamma_5 d_{k}(y) \, ,
\end{eqnarray}
 \begin{eqnarray}
  J_P(0)&=&\varepsilon^{ila} \varepsilon^{ijk}\varepsilon^{lmn}  u^T_j(0) C\gamma_5 d_k(0)\,u^T_m(0) C\gamma_5 c_n(0)\,  C\bar{c}^{T}_{a}(0) \, ,
\end{eqnarray}
the $a$, $i$, $j$, $\cdots$  are color indices. We choose the quark currents $J_{\eta_c}(x)$, $J_{\bar{D}^0}(x)$,
$J_{D^-}(x)$, $J_\mu(x)$,
$J_{\Lambda_c^+}(y)$, $J_{\Sigma_c^{++}}(y)$, $ J_{\Sigma_c^+}(y)$, $J_N(y)$ and  $ J_P(0)$ to interpolate the hadrons $\eta_c$, $\bar{D}^0$,
$D^-$, $J/\psi$, $\Lambda_c^+$, $\Sigma_c^{++}$, $ \Sigma_c^+$, $p$ and  $P_c$, respectively. Henceforth we will write  the proton  as $N$ instead of $p$ to avoid confusing  with the four momentum $p_\mu$.

 At the hadron  side, we  insert  a complete set  of intermediate  hadron  states with the
same quantum numbers as the current operators $J_{\eta_c}(x)$, $J_{\bar{D}^0}(x)$,
$J_{D^-}(x)$, $J_\mu(x)$,
$J_{\Lambda_c^+}(y)$, $ J_{\Sigma_c^+}(y)$, $J_{\Sigma_c^{++}}(y)$,  $J_N(y)$ and  $\bar{J}_P(0)$  into the correlation functions $\Pi(p,q)$ and $\Pi_{\mu}(p,q)$ respectively to obtain the hadronic representation \cite{SVZ79,PRT85}, then we isolate all the  ground  state contributions and write them down  explicitly,
\begin{eqnarray}
 \Pi_{P\eta_cN}(p,q)   & = & \frac{f_{\eta_c}m_{\eta_c}^2\lambda_P\lambda_N}{2m_c}\frac{-i \left(\!\not\!{q}+m_N \right)\left( \!\not\!{p}^\prime+m_P \right)}{\left(m_P^2-p^{\prime2}\right)\left(m_{\eta_c}^2-p^{2}\right)\left(m_{N}^2-q^{2}\right)}G_{P\eta_cN}+\cdots\,  ,
\end{eqnarray}

\begin{eqnarray}
 \Pi_{P\bar{D}^0\Lambda_c^+}(p,q)   & = & \frac{f_{D}m_{D}^2\lambda_P\lambda_{\Lambda_c}}{m_c}\frac{-i \left(\!\not\!{q}+m_{\Lambda_c} \right)\left( \!\not\!{p}^\prime+m_P \right)}{\left(m_P^2-p^{\prime2}\right)\left(m_{D}^2-p^{2}\right)\left(m_{\Lambda_c}^2-q^{2}\right)}G_{P\bar{D}^0\Lambda_c^+}+\cdots\,  ,
\end{eqnarray}

\begin{eqnarray}
 \Pi_{P\bar{D}^0\Sigma_c^+}(p,q)   & = & \frac{f_{D}m_{D}^2\lambda_P\lambda_{\Sigma^+_c}}{m_c}\frac{-i \left(\!\not\!{q}+m_{\Sigma_c} \right)\left( \!\not\!{p}^\prime+m_P \right)}{\left(m_P^2-p^{\prime2}\right)\left(m_{D}^2-p^{2}\right)\left(m_{\Sigma_c}^2-q^{2}\right)}G_{P\bar{D}^0\Sigma_c^+}+\cdots\,  ,
\end{eqnarray}

\begin{eqnarray}
 \Pi_{P D^{-}\Sigma_c^{++}}(p,q)   & = & \frac{f_{D}m_{D}^2\lambda_P\lambda_{\Sigma^{++}_c}}{m_c}\frac{-i \left(\!\not\!{q}+m_{\Sigma_c} \right)\left( \!\not\!{p}^\prime+m_P \right)}{\left(m_P^2-p^{\prime2}\right)\left(m_{D}^2-p^{2}\right)\left(m_{\Sigma_c}^2-q^{2}\right)}G_{P D^{-}\Sigma_c^{++}}+\cdots\,  ,
\end{eqnarray}

\begin{eqnarray}
 \Pi_\mu(p,q)  & = & f_{J/\psi}m_{J/\psi}\lambda_P\lambda_N\frac{- \left(\!\not\!{q}+m_N \right)\left( G_V\gamma^\alpha-i\frac{G_T}{m_P+m_N}\sigma^{\alpha\beta}p_\beta\right)\gamma_5\left( \!\not\!{p}^\prime+m_P \right)}{\left(m_P^2-p^{\prime2}\right)\left(m_{J/\psi}^2-p^{2}\right)\left(m_{N}^2-q^{2}\right)}\nonumber\\
  &&\left(-g_{\mu\alpha}+\frac{p_{\mu}p_{\alpha}}{p^2} \right)+\cdots\,  ,
\end{eqnarray}
where we introduce the subscripts $P\eta_cN$, $P\bar{D}^0\Lambda_c^+$, $P\bar{D}^0\Sigma_c^+$ and $P D^{-}\Sigma_c^{++}$ in the correlation functions $\Pi(p,q)$ to distinguish the  corresponding hadronic coupling constants, and we   take  the standard  definitions for the pole residues or decay constants $\lambda_{P}$, $\lambda_{N}$, $\lambda_{\Lambda_c}$, $\lambda_{\Sigma_c}$, $f_{\eta_c}$, $f_{D}$, $f_{J/\psi}$,
\begin{eqnarray}
\langle 0| J (0)|P_{c}(p^\prime)\rangle &=&\lambda_{P} U(p^\prime,s) \, ,  \nonumber\\
\langle 0| J_{N} (0)|N(q)\rangle &=&\lambda_{N} U(q,s) \, , \nonumber \\
\langle 0| J_{\Lambda_c} (0)|\Lambda_c(q)\rangle &=&\lambda_{\Lambda_c} U(q,s) \, , \nonumber \\
\langle 0| J_{\Sigma_c} (0)|\Sigma_c(q)\rangle &=&\lambda_{\Sigma_c} U(q,s) \, , \nonumber \\
\langle 0| J_{\eta_c} (0)|\eta_c(p)\rangle &=&\frac{f_{\eta_c}m_{\eta_c}^2}{2m_c} \, ,\nonumber \\
\langle 0| J_{D} (0)|D(p)\rangle &=&\frac{f_{D}m_{D}^2}{m_c} \, ,\nonumber \\
\langle 0| J_{\mu} (0)|J/\psi(p)\rangle &=&f_{J/\psi}m_{J/\psi}\varepsilon_\mu(p,s) \, ,
\end{eqnarray}
and the hadronic coupling constants $G_{P\eta_cN}$, $G_{P\bar{D}^0\Lambda_c^+}$, $G_{P\bar{D}^0\Sigma_c^+}$, $G_{PD^-\Sigma_c^{++}}$,  $G_V$ and $G_T$,
\begin{eqnarray}\label{Coupling-GVGT}
\langle\eta_c(p)N(q)|P_c(p^\prime)\rangle&=& G_{P\eta_cN}\,\overline{U}(q)U(p^\prime)\, ,\nonumber\\
\langle \bar{D}^0(p)\Lambda_c^+(q)|P_c(p^\prime)\rangle&=& G_{P\bar{D}^0\Lambda_c^+}\,\overline{U}(q)U(p^\prime)\, ,\nonumber\\
\langle \bar{D}^0(p)\Sigma_c^+(q)|P_c(p^\prime)\rangle&=& G_{P\bar{D}^0\Sigma_c^+}\,\overline{U}(q)U(p^\prime)\, ,\nonumber\\
\langle D^-(p)\Sigma_c^{++}(q)|P_c(p^\prime)\rangle&=& G_{PD^-\Sigma_c^{++}}\,\overline{U}(q)U(p^\prime)\, ,\nonumber\\
\langle J/\psi(p)N(q)|P_c(p^\prime)\rangle&=& -i\overline{U}(q)\varepsilon^*_\alpha\left( G_V\gamma^\alpha-i\frac{G_T}{m_P+m_N}\sigma^{\alpha\beta}p_\beta\right)\gamma_5U(p^\prime)\, ,
\end{eqnarray}
 the $U(p^\prime,s)$, $U(p,s)$ and $U(q,s)$ are the Dirac spinors, and the $\varepsilon_\mu$ is the polarization vector of the $J/\psi$.

In this article, we choose $\Gamma^\prime=\sigma_{\mu\nu}$, $\gamma_5\!\not\!{z}$, $\gamma_5$ in Eq.\eqref{Tr-Gamma}, and accomplish   the traces in
the Dirac spinor space,
\begin{eqnarray}
\frac{1}{4}{\rm Tr}\left[\Pi_{H}(p,q)\sigma_{\mu\nu}\right]&=& \Pi_H(p^{\prime2},p^2,q^2) \,\left(p_\mu q_\nu-q_\mu p_\nu \right)+\cdots\, , \nonumber\\
\frac{1}{4}{\rm Tr}\left[\Pi_\mu^{H}(p,q)\gamma_5\!\not\!{z} \right]&=& \Pi_H^1(p^{\prime2},p^2,q^2)\, q_\mu p\cdot z+\cdots\, ,\nonumber\\
\frac{1}{4}{\rm Tr}\left[\Pi_\mu^{H}(p,q)\gamma_5  \right]&=& \Pi_H^2(p^{\prime2},p^2,q^2)\, q_\mu  +\cdots\, ,
\end{eqnarray}
 and choose the tensor structures $p_\mu q_\nu-q_\mu p_\nu$,  $q_\mu p\cdot z$ and $q_\mu$ to study the hadronic coupling constants, where the $z_\mu$ is an arbitrary four-vector we introduce to select the pertinent Dirac structures. We neglect the explicit expressions of the correlation functions $\Pi_H(p^{\prime2},p^2,q^2) $, $\Pi_H^1(p^{\prime2},p^2,q^2)$ and $\Pi_H^2(p^{\prime2},p^2,q^2) $ at the hadron side for simplicity.

\begin{figure}
 \centering
  \includegraphics[totalheight=4cm,width=12cm]{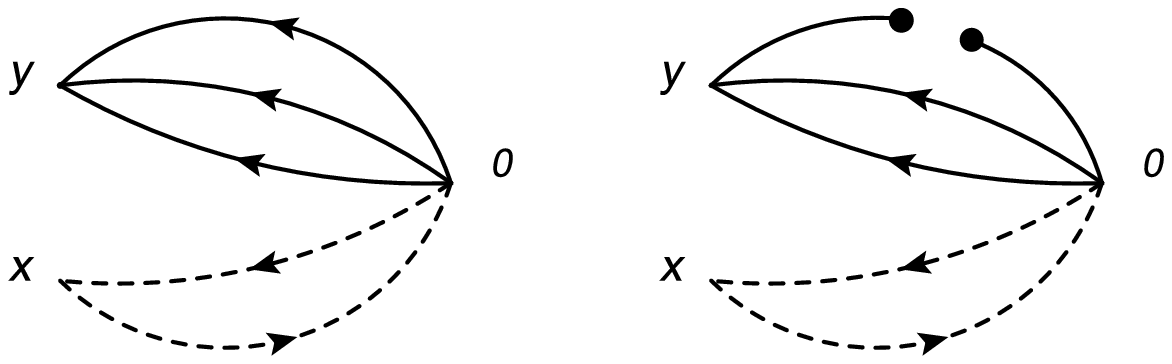} \\
  \vspace{1cm}
   \includegraphics[totalheight=4cm,width=12cm]{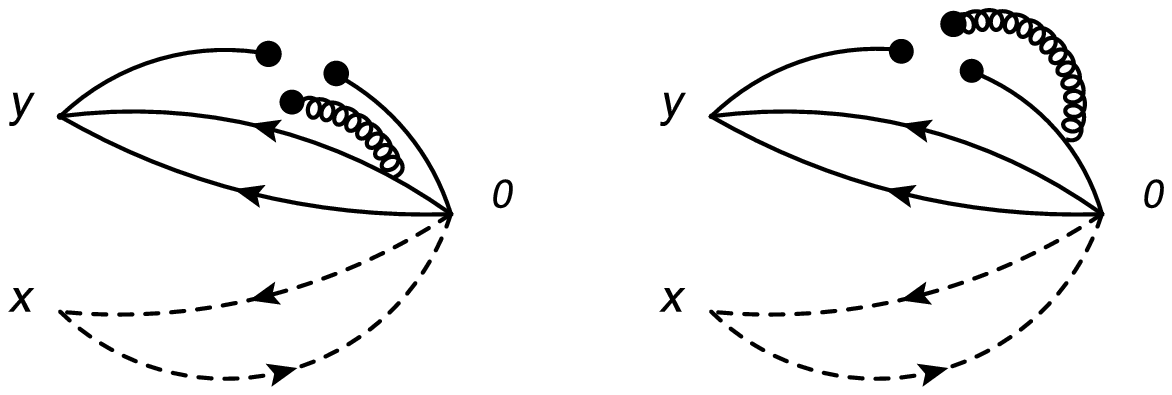}\\
    \vspace{1cm}
   \includegraphics[totalheight=4cm,width=12cm]{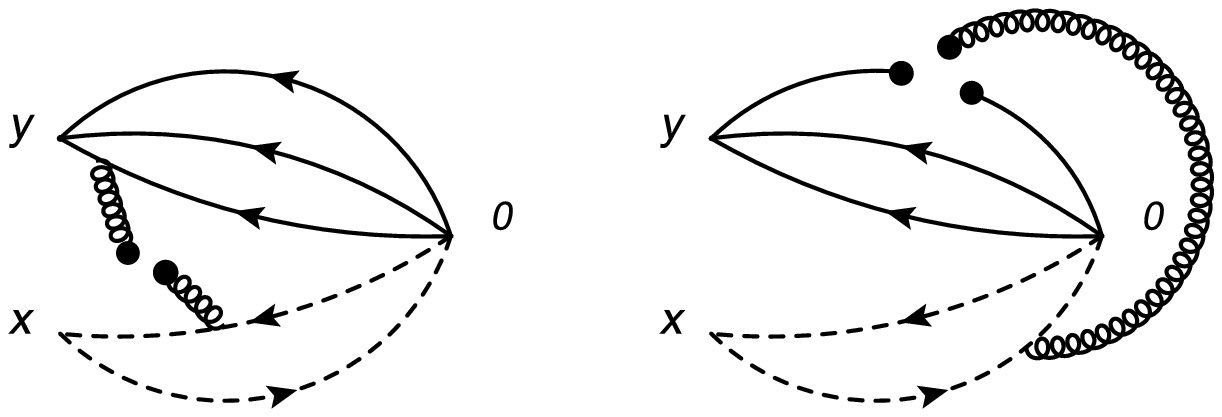}\\
    \vspace{1cm}
   \includegraphics[totalheight=4cm,width=12cm]{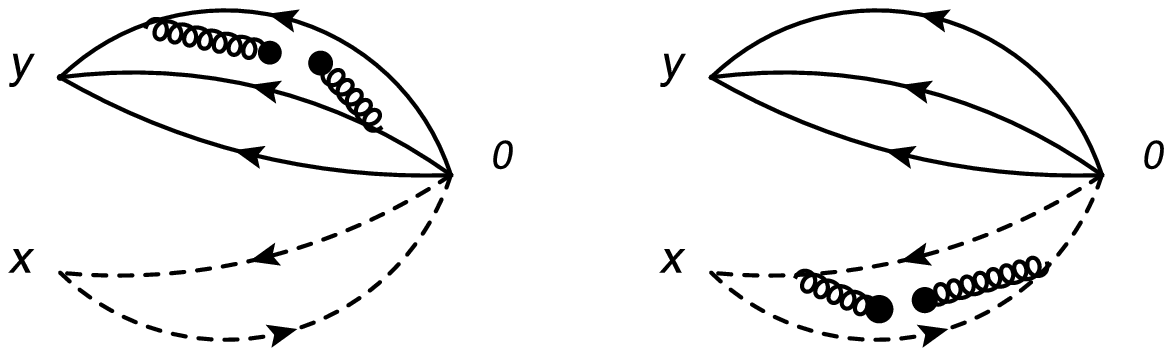}
 \caption{ The Feynman diagrams contributing to the two-body strong decays. Other
diagrams obtained by interchanging of the  light quark lines (solid lines)  or heavy quark lines (dashed lines) are implied.}\label{decay-Feynman-Diagram}
\end{figure}

At the QCD side of the correlation functions, we carry out the operator product expansion up to the vacuum condensates  of dimension-10, the interval of the vacuum condensates  is large enough to obtain stable QCD sum rules in case of single Borel parameter, the relevant Feynman diagrams are shown explicitly in Fig.\ref{decay-Feynman-Diagram}. Moreover, we  assume  vacuum saturation for the  higher dimensional  vacuum condensates. As the vacuum condensates are vacuum expectations of the quark-gluon operators of the dimensions  $n$, we take
the truncations $\mathcal{O}( \alpha_s^{k})$ with $n\leq 10$ and $k\leq 1$ in a consistent way, and write  (components of) the correlation functions  $\Pi_{QCD}(p^{\prime 2},p^2,q^2)$  as
\begin{eqnarray}
\Pi_{QCD}(p^{\prime 2},p^2,q^2)&=&  \int_{\Delta_s^2}^{s^0_{M}}ds \int_{\Delta_u^2}^{s^0_{B}}du  \frac{\rho_{QCD}(s,u)}{(s-p^2)(u-q^2)}+\cdots\, , \
\end{eqnarray}
through double dispersion relation, where the $\Pi_{QCD}(p^{\prime 2},p^2,q^2)$ represent the corresponding correlation functions of the  $\Pi_H(p^{\prime2},p^2,q^2) $, $\Pi_H^1(p^{\prime2},p^2,q^2)$ and $\Pi_H^2(p^{\prime2},p^2,q^2) $ at the QCD side  collectively for simplicity.

Here we take a short digression to discuss the vacuum saturation in performing the operator product expansion.  In the original works, Shifman,  Vainshtein and  Zakharov took
 the factorization hypothesis for the higher dimensional vacuum condensates according to  two reasons \cite{SVZ79}.  One is the rather large value of the quark condensate $\langle\bar{q}q\rangle$, the other is the duality between the quark and physical states, which implies that counting both the quark and
physical states may well become a double counting since they reproduce each other \cite{SVZ79}.

In the QCD sum rules for the traditional mesons, we always  introduce a parameter $\kappa$ to parameterize the  deviation from the factorization hypothesis by hand, for example, in the case of the four quark condensate,  $\langle\bar{q}q\rangle^2 \to \kappa\langle\bar{q}q\rangle^2$ \cite{Narison-rho,Narison-rho-2,Narison-rho-3}.  As the $\langle\bar{q}q\rangle^2$ is always companied with the fine-structure constant $\alpha_s=\frac{g_s^2}{4\pi}$, and plays a minor important role,
the deviation from $\kappa=1$, for example, $\kappa=2\sim 3$, cannot make much difference, though the value $\kappa>1$ can lead to better QCD sum rules in some cases.
In fact, the  vacuum saturation works well in the large $N_c$ limit \cite{Novikov--shifman}.

On  the contrary, in the QCD sum rules for the tetraquark, pentaquark and hexaquark (or molecular) states, the four-quark condensate plays an important role,
a large value, for example, $\kappa=2$, can destroy the platforms in the QCD sum rules for the current $J_{c\bar{c}}(x)$ in Ref.\cite{WZG-DvDvDv}. Furthermore, in calculations, we observe that the optimal value is $\kappa=1$, the vacuum saturation works well in the QCD sum rules for the multiquark states.

Up to now, all the multiquark states are studied with the  QCD sum rules  by assuming the vacuum saturation for the higher dimensional vacuum condensates tacitly in performing the operator product expansion, except for in some case the parameter $\kappa$ is introduced
for the sake of  fine-tuning. The true values (also the next-to-leading-order perturbative corrections) of the higher dimensional vacuum condensates, even the four quark condensates $\langle \bar{q}\Gamma q \bar{q}\Gamma^\prime q\rangle$, remain unknown or poorly known,  where the $\Gamma$ and $\Gamma^\prime$ stand for the Dirac $\gamma$-matrixes, we cannot obtain robust estimations  about the effects beyond the vacuum saturation.

Now we come back to the correlation functions $\Pi(p^{\prime 2},p^2,q^2)$, and  accomplish  the integral over $ds^\prime$  firstly at the hadron side  according to Eqs.\eqref{duality}-\eqref{quark-hadron-duality}, then match the hadron side  with the QCD side of the correlation functions $\Pi(p^{\prime 2},p^2,q^2)$ to obtain the rigorous   duality,
then  write down  the quark-hadron duality explicitly,
 \begin{eqnarray}
  \int_{4m_c^2}^{s^0_{\eta_c}}ds \int_{0}^{s^0_{N}}du  \frac{\rho^{\eta_cN}_{QCD}(s,u)}{(s-p^2)(u-q^2)}
  &=&\frac{f_{\eta_c}m_{\eta_c}^2\lambda_P\lambda_N}{2m_c}\frac{G_{P\eta_cN}}{\left(m_P^2-p^{\prime2}\right)\left(m_{\eta_c}^2-p^{2}\right)\left(m_{N}^2-q^{2}\right)}\nonumber\\
  &&+\frac{C_{P^{\prime}\eta_c}+C_{P^{\prime}N}}{(m_{\eta_c}^2-p^{2})(m_{N}^2-q^2)} \, ,
  \end{eqnarray}

 \begin{eqnarray}
  \int_{m_c^2}^{s^0_{D}}ds \int_{m_c^2}^{s^0_{\Lambda_c}}du  \frac{\rho^{\bar{D}^0\Lambda^+_c}_{QCD}(s,u)}{(s-p^2)(u-q^2)}
  &=&\frac{f_{D}m_{D}^2\lambda_P\lambda_{\Lambda_c}}{m_c}\frac{G_{P\bar{D}^0\Lambda^+_c}}{\left(m_P^2-p^{\prime2}\right)\left(m_{D}^2-p^{2}\right)\left(m_{\Lambda_c}^2-q^{2}\right)}\nonumber\\
  &&+\frac{C_{P^{\prime}\bar{D}^0}+C_{P^{\prime}\Lambda^+_c}}{(m_{D}^2-p^{2})(m_{\Lambda_c}^2-q^2)} \, ,
  \end{eqnarray}

\begin{eqnarray}
  \int_{m_c^2}^{s^0_{D}}ds \int_{m_c^2}^{s^0_{\Sigma_c}}du  \frac{\rho^{\bar{D}^0\Sigma^+_c}_{QCD}(s,u)}{(s-p^2)(u-q^2)}
  &=&\frac{f_{D}m_{D}^2\lambda_P\lambda_{\Sigma^+_c}}{m_c}\frac{G_{P\bar{D}^0\Sigma^+_c}}{\left(m_P^2-p^{\prime2}\right)\left(m_{D}^2-p^{2}\right)\left(m_{\Sigma_c}^2-q^{2}\right)}\nonumber\\
  &&+\frac{C_{P^{\prime}\bar{D}^0}+C_{P^{\prime}\Sigma^{+}_c}}{(m_{D}^2-p^{2})(m_{\Sigma_c}^2-q^2)} \, ,
  \end{eqnarray}

 \begin{eqnarray}
  \int_{m_c^2}^{s^0_{D}}ds \int_{m_c^2}^{s^0_{\Sigma_c}}du  \frac{\rho^{D^{-}\Sigma^{++}_c}_{QCD}(s,u)}{(s-p^2)(u-q^2)}
  &=&\frac{f_{D}m_{D}^2\lambda_P\lambda_{\Sigma^{++}_c}}{m_c}\frac{G_{PD^{-}\Sigma^{++}_c}}{\left(m_P^2-p^{\prime2}\right)\left(m_{D}^2-p^{2}\right)\left(m_{\Sigma_c}^2-q^{2}\right)}\nonumber\\
  &&+\frac{C_{P^{\prime}D^{-}}+C_{P^{\prime}\Sigma^{++}_c}}{(m_{D}^2-p^{2})(m_{\Sigma_c}^2-q^2)} \, ,
  \end{eqnarray}

  \begin{eqnarray}
  \int_{4m_c^2}^{s^0_{J/\psi}}ds \int_{0}^{s^0_{N}}du  \frac{\rho^{J/\psi N,1}_{QCD}(s,u)}{(s-p^2)(u-q^2)}
  &=&f_{J/\psi}m_{J/\psi}\lambda_P\lambda_N \frac{G_T-G_V}{\left(m_P^2-p^{\prime2}\right)\left(m_{J/\psi}^2-p^{2}\right)\left(m_{N}^2-q^{2}\right)}\nonumber\\
  &&+\frac{C_{P^{\prime}J/\psi,1}+C_{P^{\prime}N,1}}{(m_{J/\psi}^2-p^{2})(m_{N}^2-q^2)} \, ,
\end{eqnarray}

 \begin{eqnarray}
  \int_{4m_c^2}^{s^0_{J/\psi}}ds \int_{0}^{s^0_{N}}du  \frac{\rho^{J/\psi N,2}_{QCD}(s,u)}{(s-p^2)(u-q^2)}
  &=&f_{J/\psi}m_{J/\psi}\lambda_P\lambda_N \frac{\left(m_P-m_N\right)G_V-G_T\frac{m_{J/\psi}^2}{m_P+m_N}}{\left(m_P^2-p^{\prime2}\right)\left(m_{J/\psi}^2-p^{2}\right)\left(m_{N}^2-q^{2}\right)}\nonumber\\
  &&+\frac{C_{P^{\prime}J/\psi,2}+C_{P^{\prime}N,2}}{(m_{J/\psi}^2-p^{2})(m_{N}^2-q^2)} \, ,
\end{eqnarray}
where the parameters $C_{P^{\prime}\eta_c}+C_{P^{\prime}N}$, $C_{P^{\prime}\bar{D}^0}+C_{P^{\prime}\Lambda^+_c}$, $C_{P^{\prime}\bar{D}^0}+C_{P^{\prime}\Sigma^{+}_c}$,
$C_{P^{\prime}D^{-}}+C_{P^{\prime}\Sigma^{++}_c}$, $C_{P^{\prime}J/\psi,1}+C_{P^{\prime}N,1}$ and $C_{P^{\prime}J/\psi,2}+C_{P^{\prime}N,2}$ are defined according to Eq.\eqref{subtract-constants}.

 We perform  double Borel transform  with respect to the variables $P^2=-p^2$ and $Q^2=-q^2$ respectively according to the routines in Eqs.\eqref{QCDSR-4P2}-\eqref{QCDSR-4Q2} to obtain the  QCD sum rules,
\begin{eqnarray}\label{QCDSR-etacN}
&&\frac{f_{\eta_c}m_{\eta_c}^2\lambda_P\lambda_N}{2m_c} \frac{G_{P\eta_cN}}{m_{P}^2-m_{\eta_c}^2} \left[ \exp\left(-\frac{m_{\eta_c}^2}{T_1^2} \right)-\exp\left(-\frac{m_{P}^2}{T_1^2} \right)\right]\exp\left(-\frac{m_{N}^2}{T_2^2} \right) +\nonumber\\
&&\left(C_{P^{\prime}\eta_c}+C_{P^{\prime}N}\right) \exp\left(-\frac{m_{\eta_c}^2}{T_1^2} -\frac{m_{N}^2}{T_2^2} \right)=\int_{4m_c^2}^{s_{\eta_c}^0} ds \int_{0}^{s_N^0} du\, \rho^{\eta_cN}_{QCD}(s,u)\exp\left(-\frac{s}{T_1^2} -\frac{u}{T_2^2} \right)\, ,\nonumber\\
\end{eqnarray}

\begin{eqnarray}\label{QCDSR-D0Lambdac}
&&\frac{f_{D}m_{D}^2\lambda_P\lambda_{\Lambda_c}}{4m_c} \frac{G_{P\bar{D}^0\Lambda_c^+}}{\widetilde{m}_{P}^2-m_{\Lambda_c}^2} \left[ \exp\left(-\frac{m_{\Lambda_c}^2}{T_2^2} \right)-\exp\left(-\frac{\widetilde{m}_{P}^2}{T_2^2} \right)\right]\exp\left(-\frac{m_{D}^2}{T_1^2} \right) +\nonumber\\
&&\left(C_{P^{\prime}\bar{D}^0}+C_{P^{\prime}\Lambda_c^+}\right) \exp\left(-\frac{m_{\Lambda_c}^2}{T_2^2} -\frac{m_{D}^2}{T_1^2} \right)=\int_{m_c^2}^{s_{D}^0} ds \int_{m_c^2}^{s_{\Lambda_c}^0} du\, \rho^{\bar{D}^0\Lambda_c^+}_{QCD}(s,u)\exp\left(-\frac{s}{T_1^2} -\frac{u}{T_2^2} \right)\, ,\nonumber\\
\end{eqnarray}

\begin{eqnarray}\label{QCDSR-D0Sigmac}
&&\frac{f_{D}m_{D}^2\lambda_P\lambda_{\Sigma^{+}_c}}{4m_c} \frac{G_{P\bar{D}^0\Sigma_c^+}}{\widetilde{m}_{P}^2-m_{\Sigma_c}^2} \left[ \exp\left(-\frac{m_{\Sigma_c}^2}{T_2^2} \right)-\exp\left(-\frac{\widetilde{m}_{P}^2}{T_2^2} \right)\right]\exp\left(-\frac{m_{D}^2}{T_1^2} \right) +\nonumber\\
&&\left(C_{P^{\prime}\bar{D}^0}+C_{P^{\prime}\Sigma_c^+}\right) \exp\left(-\frac{m_{\Sigma_c}^2}{T_2^2} -\frac{m_{D}^2}{T_1^2} \right)=\int_{m_c^2}^{s_{D}^0} ds \int_{m_c^2}^{s_{\Sigma_c}^0} du\, \rho^{\bar{D}^0\Sigma_c^+}_{QCD}(s,u)\exp\left(-\frac{s}{T_1^2} -\frac{u}{T_2^2} \right)\, ,\nonumber\\
\end{eqnarray}

\begin{eqnarray}\label{QCDSR-D0SigmacPP}
&&\frac{f_{D}m_{D}^2\lambda_P\lambda_{\Sigma^{++}_c}}{4m_c} \frac{G_{PD^{-}\Sigma_c^{++}}}{\widetilde{m}_{P}^2-m_{\Sigma_c}^2} \left[ \exp\left(-\frac{m_{\Sigma_c}^2}{T_2^2} \right)-\exp\left(-\frac{\widetilde{m}_{P}^2}{T_2^2} \right)\right]\exp\left(-\frac{m_{D}^2}{T_1^2} \right) +\nonumber\\
&&\left(C_{P^{\prime}D^{-}}+C_{P^{\prime}\Sigma_c^{++}}\right) \exp\left(-\frac{m_{\Sigma_c}^2}{T_2^2} -\frac{m_{D}^2}{T_1^2} \right)=\int_{m_c^2}^{s_{D}^0} ds \int_{m_c^2}^{s_{\Sigma_c}^0} du\, \rho^{D^{-}\Sigma_c^{++}}_{QCD}(s,u)\exp\left(-\frac{s}{T_1^2} -\frac{u}{T_2^2} \right)\, ,\nonumber\\
\end{eqnarray}

\begin{eqnarray}\label{QCDSR-GV-GT}
&&f_{J/\psi}m_{J/\psi}\lambda_P\lambda_N \frac{G_{V/T}}{m_{P}^2-m_{J/\psi}^2} \left[ \exp\left(-\frac{m_{J/\psi}^2}{T_1^2} \right)-\exp\left(-\frac{m_{P}^2}{T_1^2} \right)\right]\exp\left(-\frac{m_{N}^2}{T_2^2} \right) +\nonumber\\
&&C_{V/T} \exp\left(-\frac{m_{J/\psi}^2}{T_1^2} -\frac{m_{N}^2}{T_2^2} \right)=\int_{4m_c^2}^{s_{J/\psi}^0} ds \int_{0}^{s_N^0} du\, \rho^{V/T}_{QCD}(s,u)\exp\left(-\frac{s}{T_1^2} -\frac{u}{T_2^2} \right)\, ,
\end{eqnarray}

\begin{eqnarray}
C_V&=&\left[\frac{m_{J/\psi}^2}{m_P+m_N}\left(C_{P^{\prime} J/\psi,1}+C_{P^{\prime} N,1}\right)+\left(C_{P^{\prime} J/\psi,2}+C_{P^{\prime} N,2}\right) \right]\frac{m_P+m_N}{m_P^2-m_{J/\psi}^2-m_N^2}\, ,\nonumber\\
C_T&=&\left[\left(m_P-m_N\right)\left(C_{P^{\prime} J/\psi,1}+C_{P^{\prime} N,1}\right)+\left(C_{P^{\prime} J/\psi,2}+C_{P^{\prime} N,2}\right) \right]\frac{m_P+m_N}{m_P^2-m_{J/\psi}^2-m_N^2}\, ,\nonumber\\
\rho_{QCD}^V(s,u)&=&\left[\frac{m_{J/\psi}^2}{m_P+m_N}\rho_{QCD}^{J/\psi N,1}(s,u)+\rho_{QCD}^{J/\psi N,2}(s,u) \right]\frac{m_P+m_N}{m_P^2-m_{J/\psi}^2-m_N^2}\, ,\nonumber\\
\rho_{QCD}^T(s,u)&=&\left[\left(m_P-m_N\right)\rho_{QCD}^{J/\psi N,1}(s,u)+\rho_{QCD}^{J/\psi N,2}(s,u) \right]\frac{m_P+m_N}{m_P^2-m_{J/\psi}^2-m_N^2}\, ,
\end{eqnarray}
in the isospin limit, $\lambda_{\Sigma_c^{++}}=\sqrt{2}\lambda_{\Sigma_c^{+}}$,  $\rho^{D^{-}\Sigma_c^{++}}_{QCD}(s,u)=2\rho^{\bar{D}^{0}\Sigma_c^{+}}_{QCD}(s,u)$,
 then we can obtain the relation $G_{PD^{-}\Sigma_c^{++}}=\sqrt{2}G_{P\bar{D}^{0}\Sigma_c^{+}}$, and neglect the QCD sum rules in Eq.\eqref{QCDSR-D0SigmacPP} in the numerical calculations, the explicit expressions of the QCD spectral densities $\rho^{\eta_cN}_{QCD}(s,u)$, $\rho^{\bar{D}^0\Lambda_c^+}_{QCD}(s,u)$,
$\rho^{\bar{D}^0\Sigma_c^+}_{QCD}(s,u)$,  $\rho^{J/\psi N,1}_{QCD}(s,u)$ and $\rho^{J/\psi N,2}_{QCD}(s,u)$ are given in the Appendix.
   Moreover,  we set the two Borel parameters to be $T_1^2=T_2^2=T^2$ according to the arguments after Eq.\eqref{QCDSR-4Q2}.

\section{Numerical results and discussions}
At the hadron side, we take the hadronic parameters  as
$m_{J/\psi}=3.0969\,\rm{GeV}$, $m_{N}=0.93827\,\rm{GeV}$,
$m_{\eta_c}=2.9839\,\rm{GeV}$,  $m_{\bar{D}^0}=1.86484\,\rm{GeV}$, $m_{\Lambda_c}=2.28646\,\rm{GeV}$, $m_{\Sigma_c}=2.4529\,\rm{GeV}$ from the Particle Data Group \cite{PDG}, and $\sqrt{s^0_{J/\psi}}=3.6\,\rm{GeV}$, $\sqrt{s^0_{\eta_c}}=3.5\,\rm{GeV}$, $\sqrt{s^0_{N}}=1.3\,\rm{GeV}$,
$f_{J/\psi}=0.418 \,\rm{GeV}$, $f_{\eta_c}=0.387 \,\rm{GeV}$  \cite{Becirevic}, $\sqrt{s^0_{D}}=2.5\,\rm{GeV}$,  $f_{D}=0.208 \,\rm{GeV}$  \cite{WangZG-decay-cnstant-D},
$\lambda_N=0.032\,\rm{GeV}^3$ \cite{SB-Ioffe}, $\sqrt{s^0_{\Lambda_c}}=3.1\,\rm{GeV}$,  $\lambda_{\Lambda_c}=0.022\,\rm{GeV}^3$ \cite{WangZG-Lambdac}, $\sqrt{s^0_{\Sigma_c}}=3.2\,\rm{GeV}$, $\lambda_{\Sigma_c^{+}}=0.045\,\rm{GeV}^3$ \cite{WangZG-Sigmac}, $m_{P}=4.31\,\rm{GeV}$, $\lambda_P=1.40\times 10^{-3}\,\rm{GeV}^6$  \cite{WangZG-penta-2020-IJMPA} from the QCD sum rules.

At the QCD side, we take  the standard values of the vacuum condensates $\langle
\bar{q}q \rangle=-(0.24\pm 0.01\, \rm{GeV})^3$,   $\langle
\bar{q}g_s\sigma G q \rangle=m_0^2\langle \bar{q}q \rangle$,
$m_0^2=(0.8 \pm 0.1)\,\rm{GeV}^2$,   $\langle \frac{\alpha_s
GG}{\pi}\rangle=(0.33\,\rm{GeV})^4 $    at the energy scale  $\mu=1\, \rm{GeV}$
\cite{PRT85,SVZ79,ColangeloReview}, and choose the $\overline{MS}$ mass  $m_{c}(m_c)=(1.275\pm0.025)\,\rm{GeV}$
 from the Particle Data Group \cite{PDG}.
Moreover, we take into account the energy-scale dependence of  the  parameters from the re-normalization group equation,
\begin{eqnarray}
\langle\bar{q}q \rangle(\mu)&=&\langle\bar{q}q \rangle({\rm 1 GeV})\left[\frac{\alpha_{s}({\rm 1 GeV})}{\alpha_{s}(\mu)}\right]^{\frac{12}{25}}\, , \nonumber\\
 \langle\bar{q}g_s \sigma Gq \rangle(\mu)&=&\langle\bar{q}g_s \sigma Gq \rangle({\rm 1 GeV})\left[\frac{\alpha_{s}({\rm 1 GeV})}{\alpha_{s}(\mu)}\right]^{\frac{2}{25}}\, , \nonumber\\
m_c(\mu)&=&m_c(m_c)\left[\frac{\alpha_{s}(\mu)}{\alpha_{s}(m_c)}\right]^{\frac{12}{25}} \, ,\nonumber\\
\alpha_s(\mu)&=&\frac{1}{b_0t}\left[1-\frac{b_1}{b_0^2}\frac{\log t}{t} +\frac{b_1^2(\log^2{t}-\log{t}-1)+b_0b_2}{b_0^4t^2}\right]\, ,
\end{eqnarray}
   where $t=\log \frac{\mu^2}{\Lambda^2}$, $b_0=\frac{33-2n_f}{12\pi}$, $b_1=\frac{153-19n_f}{24\pi^2}$,
   $b_2=\frac{2857-\frac{5033}{9}n_f+\frac{325}{27}n_f^2}{128\pi^3}$,
   $\Lambda=210\,\rm{MeV}$, $292\,\rm{MeV}$  and  $332\,\rm{MeV}$ for the flavors
   $n_f=5$, $4$ and $3$, respectively  \cite{PDG,Narison-mix}, and evolve all the parameters to the acceptable energy scale $\mu$  with $n_f=4$ to extract the hadronic coupling constants $G_{P\eta_cN}$, $G_{P\bar{D}^0\Lambda_c^+}$, $G_{P\bar{D}^0\Sigma_c^+}$,  $G_V$ and $G_T$, as  the hidden-charm pentaquark state, charmonium states, charmed mesons and charmed baryons are involved.

  The best energy scale of the QCD spectral density in the QCD sum rules for the lowest diquark-diquark-antiquark type  hidden-charm pentaquark state with the spin-parity $J^P={\frac{1}{2}}^-$ is $\mu=2.3\,\rm{GeV}$ \cite{Pc4312-Wang-mole-decay}, which is fixed  by the energy scale formula $\mu=\sqrt{M^2_{X/Y/Z/P}-(2{\mathbb{M}}_c)^2}$ with the effective $c$-quark mass ${\mathbb{M}}_c=1.82\,\rm{GeV}$ in the case of the constituents are the charmed diquark (antidiquark) states in the color antitriplet (triplet) \cite{Wang-tetra-formula-1,Wang-tetra-formula-2,Wang-tetra-formula-3,WangZG-eff-Mc}. The energy scale $\mu=2.3\,\rm{GeV}$ is tool large in the QCD sum rules for the mesons $\eta_c$, $\bar{D}^0$,
$D^-$, $J/\psi$ and baryons $\Lambda_c^+$, $\Sigma_c^{++}$, $ \Sigma_c^+$, $N$.
     In this article, we take the energy scales of the QCD spectral densities to be $\mu=\frac{m_{\eta_c}}{2}=1.5\,\rm{GeV}$, which is acceptable for the
    charmed mesons and  charmonium states at least based on our previous studies \cite{WangHuangTao-3900}.
    In Ref.\cite{Narison-mu}, R. Albuquerque et al try to obtain the energy scale independent QCD sum rules for the tetraquark (molecular) states, however, they choose too large continuum threshold parameters and too large energy scales of the QCD spectral densities, for example, for the $D\bar{D}$ molecular state with the quantum numbers $J^{PC}=0^{++}$, they choose the energy scale $\mu=4.5\,\rm{GeV}$ and continuum threshold parameter $\sqrt{s_0}=2M_{D}+(1.1\sim1.9)\,\rm{GeV}=3.739+(1.1\sim1.9)\,\rm{GeV}$, and obtain the mass of the   molecular state, $M_{D\bar{D}(0^{++})}=3.898\pm 0.036\,\rm{GeV}$, the contaminations from  the higher resonances and continuum states  are already included in.

In the two-point QCD sum rules for the  hidden-charm pentaquark states $P_c$, conventional baryons $B$ and mesons $M$, see Eqs.\eqref{residue-QCDSR}-\eqref{mass-QCDSR},
the predicted masses $m_{P/B/M}$ and pole residues $\lambda_{P/B/M}$ vary with the input parameters at the QCD side, such as the vacuum condensates, quark masses and continuum threshold parameters. If we choose the masses $m_{P/B/M}$ and pole residues $\lambda_{P/B/M}$ as independent parameters, which  have their
own uncertainties, we should overestimate the uncertainties. According to arguments in section 2, we neglect the uncertainties of the hadron masses and continuum threshold parameters for simplicity, and take into account the uncertainties originate from other input parameters besides the Borel parameters at the QCD side.

In the three-point QCD sum rules, we usually set $T_1^2=T_2^2=T^2$ by merging the two Borel parameters in Eqs.\eqref{QCDSR-4P2}-\eqref{QCDSR-4Q2} to obtain stable QCD sum rules,
\begin{eqnarray}\label{QCDSR-4P2-T2}
&& \frac{\lambda_{P}\lambda_{M}\lambda_{B}G_{PMB}}{m_{P}^2-m_{M}^2} \left[ \exp\left(-\tau m_{M}^2 \right)-\exp\left(-\tau m_{P}^2 \right)\right]\exp\left(-\tau m_{B}^2 \right) +\nonumber\\
&&\left(C_{P^{\prime}M}+C_{P^{\prime}B}\right) \exp\left(-\tau m_{M}^2 -\tau m_{B}^2  \right)=\int_{\Delta_s^2}^{s_M^0} ds \int_{\Delta_u^2}^{s_B^0} du\, \rho_{QCD}(s,u)\exp\left(-\tau s -\tau u \right)\, ,
\end{eqnarray}
\begin{eqnarray}\label{QCDSR-4Q2-T2}
&& \frac{\lambda_{P}\lambda_{M}\lambda_{B}G_{PMB}}{4\left(\widetilde{m}_{P}^2-m_{B}^2\right)} \left[ \exp\left(-\tau m_{B}^2 \right)-\exp\left(-\tau\widetilde{m}_{P}^2  \right)\right]\exp\left(-\tau m_{M}^2  \right) +\nonumber\\
&&\left(C_{P^{\prime}M}+C_{P^{\prime}B}\right) \exp\left(-\tau m_{B}^2  -\tau m_{M}^2  \right)=\int_{\Delta_s^2}^{s_M^0} ds \int_{\Delta_u^2}^{s_B^0} du\, \rho_{QCD}(s,u)\exp\left(-\tau s -\tau u  \right)\, ,
\end{eqnarray}
where $\tau=\frac{1}{T^2}$.
We can choose $\xi$ to stand for the vacuum condensates $\langle\bar{q}q\rangle$, $\langle\bar{q}g_s\sigma Gq\rangle$, $\langle \frac{\alpha_sGG}{\pi}\rangle$  or $c$-quark mass,  the uncertainties   $\xi \to \xi +\delta \xi$ lead to the uncertainties $\lambda_{P}\lambda_{M}\lambda_{B}G_{PMB} \to \lambda_{P}\lambda_{M}\lambda_{B}G_{PMB}+\delta\,\lambda_{P}\lambda_{M}\lambda_{B}G_{PMB}$, $C_{P^{\prime}M} \to C_{P^{\prime}M}+\delta C_{P^{\prime}M}$,
$C_{P^{\prime}B} \to C_{P^{\prime}B}+\delta C_{P^{\prime}B}$,
where
\begin{eqnarray}\label{Uncertainty-4}
\delta\,\lambda_{P}\lambda_{M}\lambda_{B}G_{PMB} &=&\lambda_{P}\lambda_{M}\lambda_{B}G_{PMB}\left( \frac{\delta \lambda_{P}}{\lambda_{P}} +\frac{\delta \lambda_{M}}{\lambda_{M}}+\frac{\delta \lambda_{B}}{\lambda_{B}}+\frac{\delta G_{PMB}}{G_{PMB}}\right)\, .
\end{eqnarray}

In the two-point QCD sum rules, we can also choose $\xi$ to stand for the vacuum condensates $\langle\bar{q}q\rangle$, $\langle\bar{q}g_s\sigma Gq\rangle$, $\langle \frac{\alpha_sGG}{\pi}\rangle$  or $c$-quark mass,  the uncertainties   $\xi \to \xi +\delta \xi$ lead to the uncertainties
$\lambda_{P} \to \lambda_{P}+\delta\,\tilde{\lambda}_{P}$, $\lambda_{B} \to \lambda_{B}+\delta\,\tilde{\lambda}_{B}$, $\lambda_{M} \to \lambda_{M}+\delta\,\tilde{\lambda}_{M}$. The uncertainties maybe have the relations $\lambda_{i}\neq \tilde{\lambda}_{i}$ or $\lambda_{i}\approx \tilde{\lambda}_{i}$ with $i=P$, $B$, $M$.  In calculations, we can take the approximation $\frac{\delta \lambda_{P}}{\lambda_{P}} =\frac{\delta \lambda_{M}}{\lambda_{M}}=\frac{\delta \lambda_{B}}{\lambda_{B}}=\frac{\delta G_{PMB}}{G_{PMB}}$ in Eq.\eqref{Uncertainty-4}, or equivalently, set $\frac{\delta \lambda_{P}}{\lambda_{P}} =\frac{\delta \lambda_{M}}{\lambda_{M}}=\frac{\delta \lambda_{B}}{\lambda_{B}}=0$, and obtain the uncertainties $\frac{\delta G_{PMB}}{G_{PMB}}$, then take the replacement,
\begin{eqnarray}\label{replace}
\frac{\delta G_{PMB}}{G_{PMB}} &\to& \frac{1}{4}\frac{\delta G_{PMB}}{G_{PMB}}\, ,
\end{eqnarray}
to avoid overestimating the uncertainties of the hadronic coupling constants.

We choose the values of the free parameters  as
$C_{P^{\prime}\eta_c}+C_{P^{\prime}N}=-1.167\times 10^{-5}\,\rm{GeV}^9$,
$C_{P^{\prime}\bar{D}^0}+C_{P^{\prime}\Lambda_c^+}=1.24\times 10^{-6}{\rm{GeV}^7} T^2$,
$C_{P^{\prime}\bar{D}^0}+C_{P^{\prime}\Sigma_c^+}=-5.14\times 10^{-6}{\rm{GeV}^7} T^2$,
$C_{V}=2.406\times 10^{-5}\,\rm{GeV}^9  $,
$C_{T}=4.12\times 10^{-6}\,{\rm{GeV}^{8}}\sqrt{T^2} $
   to obtain flat platforms in the Borel windows
   $T^2 =(4.7-5.7)\,\rm{GeV^2}$,
   $(2.3-3.1)\,\rm{GeV^2}$,
   $(1.9-2.7)\,\rm{GeV^2}$,
   $(3.5-4.5)\,\rm{GeV^2}$
    and $(3.1-4.1)\,\rm{GeV^2}$ for the hadronic coupling constants $G_{P\eta_cN}$, $G_{P\bar{D}^0\Lambda_c^+}$, $G_{P\bar{D}^0\Sigma_c^+}$, $G_V$ and $G_T$, respectively. We fit those values  to obtain the same intervals of  flat  platforms $T_{max}^2-T^2_{min}=1.0\,\rm{GeV}^2$ and $0.8\,\rm{GeV}^2$ for the hadronic coupling constants  $G_{PMB}$ in the case of  $M=$ charmonium states and $D$ mesons, respectively \cite{WangZG-solid-4660,WangZhang-Solid-1,WangZhang-Solid-2,WangZhang-Solid-3,WangZhang-Solid-4}, where the $T^2_{max}$ and $T^2_{min}$ are  the maximum and minimum values of the Borel parameters, respectively.
    As the uncertainties  $\xi \to \xi+\delta \xi$ lead  to uncertainties of the unknown functions (or free parameters) $C_{P^{\prime}M} \to C_{P^{\prime}M}+\delta C_{P^{\prime}M}$,
$C_{P^{\prime}B} \to C_{P^{\prime}B}+\delta C_{P^{\prime}B}$, we have to vary the $C_{P^{\prime}M}$ and $C_{P^{\prime}B}$ accordingly to obtain stable QCD sum rules for the hadronic coupling constants $G_{PMB}$ with variations of the Borel parameters.

In the QCD sum rules for the hadronic coupling constants among the pentaquark (or tetraquark) states and two conventional hadrons, we have to introduce the parameters $C_{P^\prime MB}$ (or $C_{X/Y/Z^\prime MM}$) to subtract the contributions from the higher resonances and continuum states in the channel $s^\prime$. In fact, we have no knowledge about the values  of the parameters  $C_{P^\prime MB}$ and $C_{X/Y/Z^\prime MM}$, even whether or not they depend  on the Borel parameters $T^2$. We resort to the same criterion in all the QCD sum rules to choose the Borel parameters $T^2$, i.e. we vary the parameters $C_{P^\prime MB}$ or $C_{X/Y/Z^\prime MM}$ via trial and error to obtain the same intervals of  flat  platforms $T_{max}^2-T^2_{min}=1.0\,\rm{GeV}^2$ and $0.8\,\rm{GeV}^2$ for the hadronic coupling constants  $G_{PMB}$ ($G_{X/Y/Z MM}$) in the case of  $M=$ charmonium states and $D$ mesons, respectively \cite{WangZG-solid-4660,WangZhang-Solid-1,WangZhang-Solid-2,WangZhang-Solid-3,WangZhang-Solid-4}, which works well in all the QCD sum rules.

\begin{figure}
\centering
\includegraphics[totalheight=6cm,width=7cm]{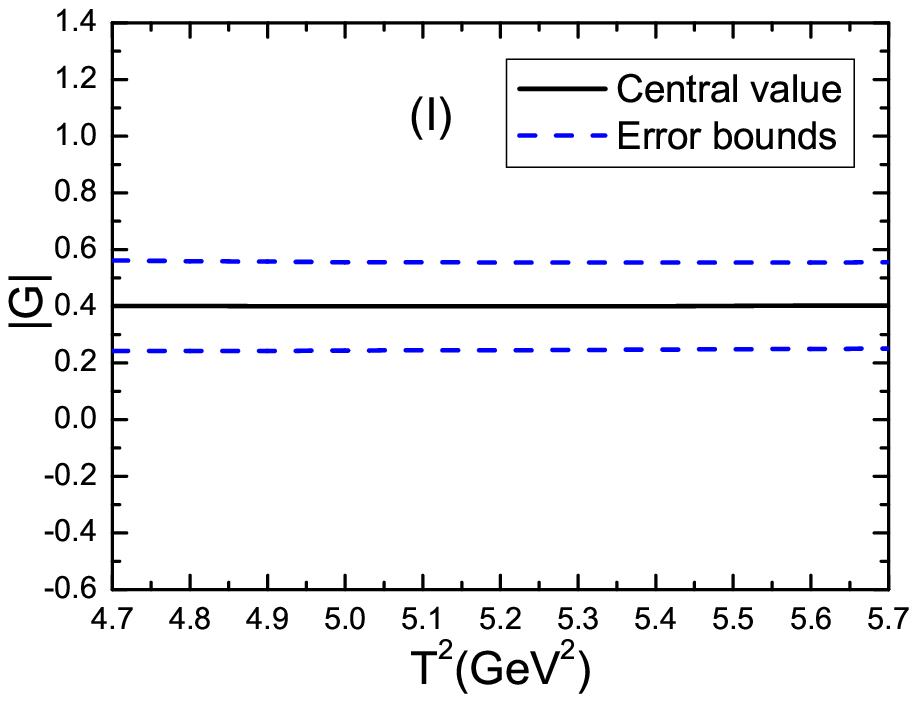}
\includegraphics[totalheight=6cm,width=7cm]{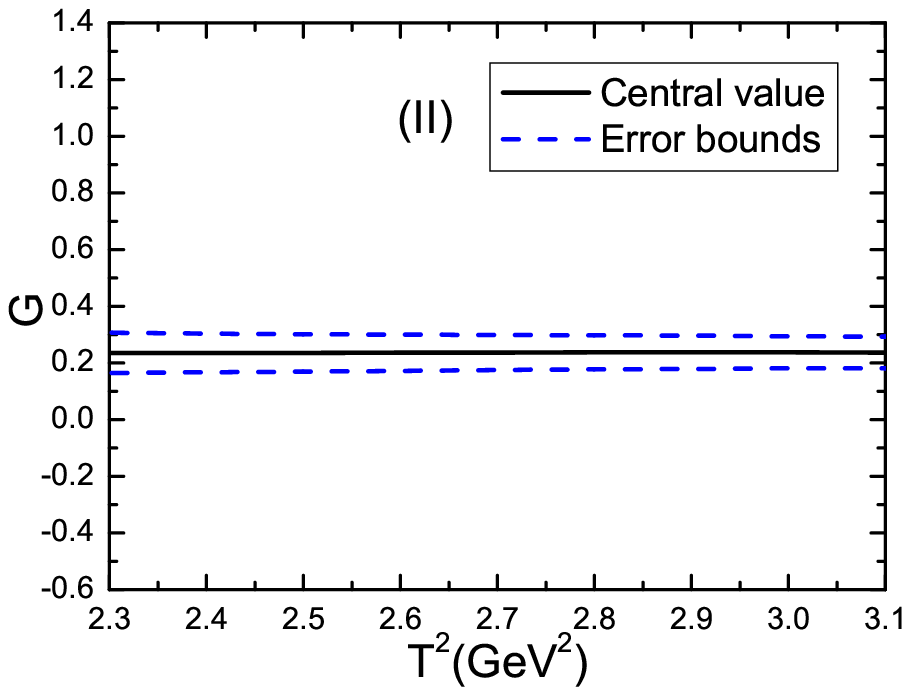}
\includegraphics[totalheight=6cm,width=7cm]{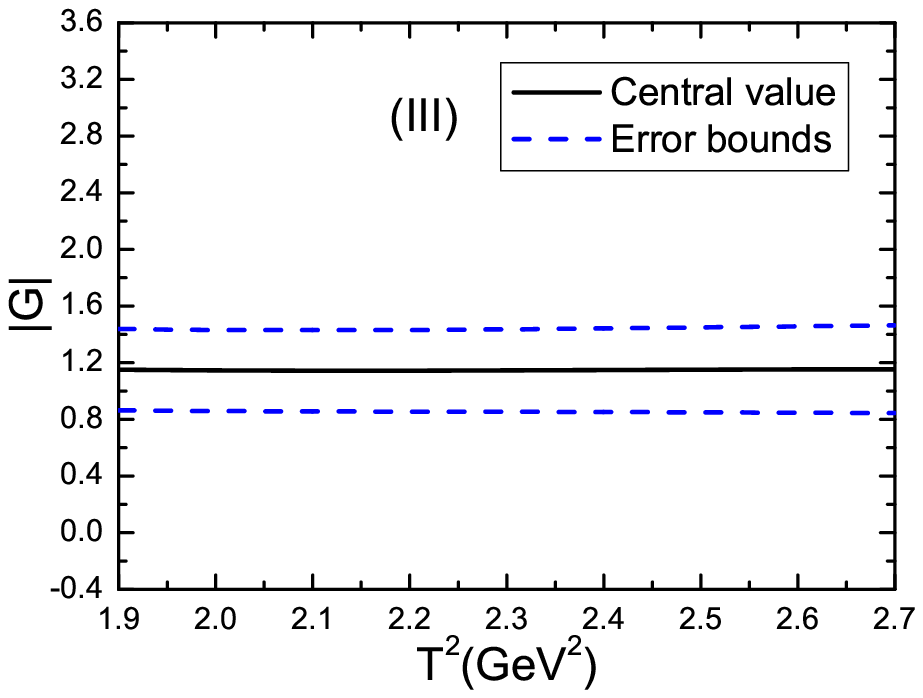}
\includegraphics[totalheight=6cm,width=7cm]{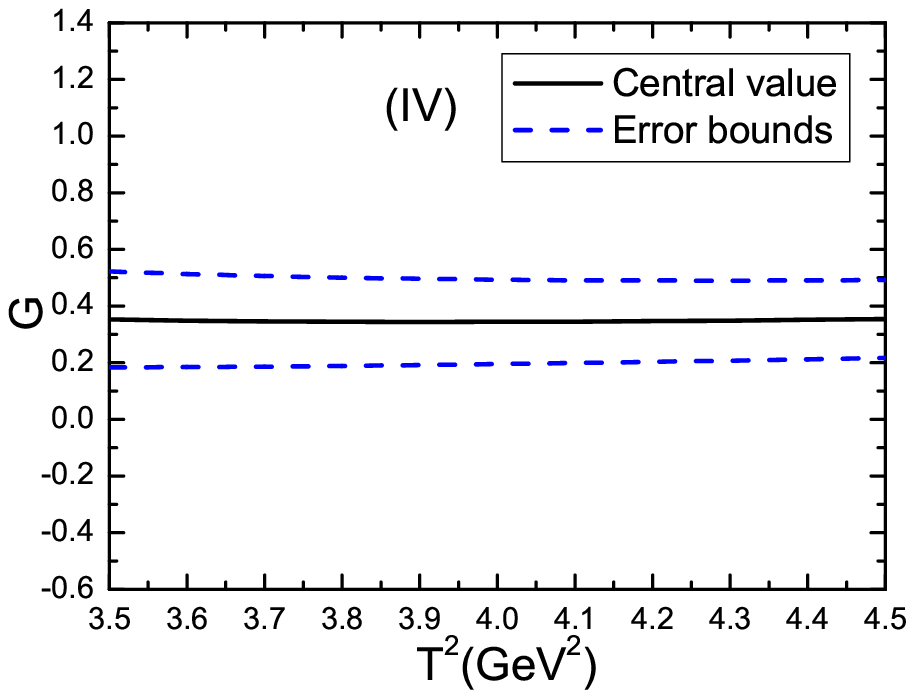}
\includegraphics[totalheight=6cm,width=7cm]{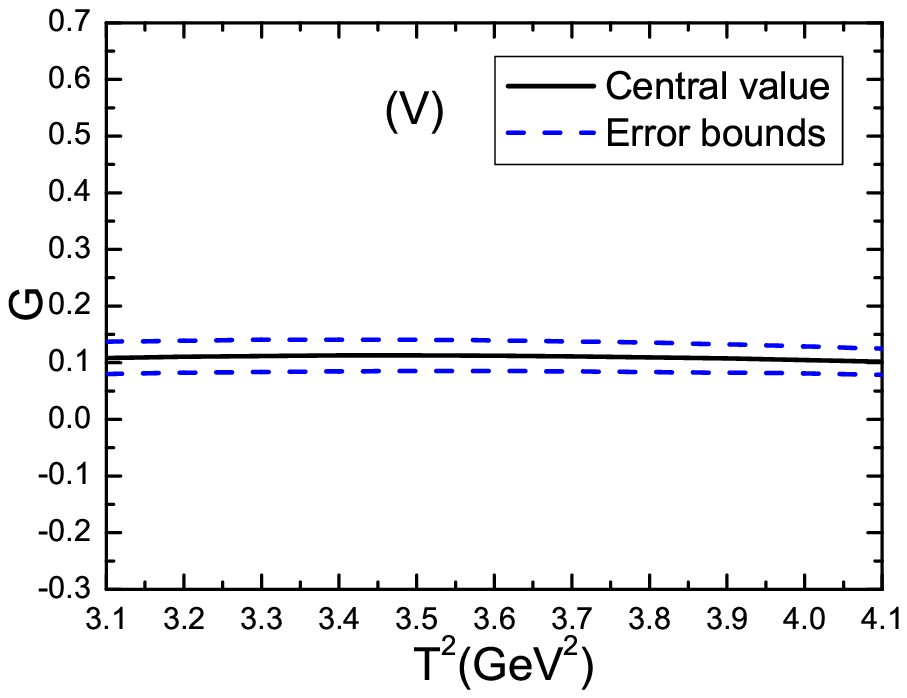}
  \caption{ The hadronic coupling constants   with variations of the  Borel parameters $T^2$, where the (I), (II), (III), (IV) and (V) correspond to the $G_{P\eta_cN}$, $G_{P\bar{D}^0\Lambda_c^+}$, $G_{P\bar{D}^0\Sigma_c^+}$, $G_V$ and $G_T$, respectively. }\label{coupling-constants-fig}
\end{figure}

Finally, we take into account  the uncertainties  of the input   parameters at the QCD side,
and obtain the values of  the hadronic coupling constants $G_{P\eta_cN}$, $G_{P\bar{D}^0\Lambda_c^+}$, $G_{P\bar{D}^0\Sigma_c^+}$, $G_V$ and $G_T$, which are shown in Fig.\ref{coupling-constants-fig},
\begin{eqnarray}
|G_{P\eta_cN}|&=&0.40\pm0.16 \nonumber\\
&\to& 0.40\pm0.04\, ,\nonumber\\
G_{P\bar{D}^0\Lambda_c^+}&=&0.24\pm0.06\nonumber\\
&\to& 0.24\pm0.02 \, ,\nonumber\\
|G_{P\bar{D}^0\Sigma_c^+}|&=&1.15\pm0.31\nonumber\\
&\to&1.15\pm0.08\, ,\nonumber\\
G_V&=&0.35\pm0.16 \nonumber\\
&\to& 0.35\pm0.04\, ,\nonumber\\
G_T&=&0.11\pm0.03 \nonumber\\
&\to& 0.11\pm0.01\, .
\end{eqnarray}

Now it is straightforward to  calculate the partial decay widths of the two-body strong decays,
\begin{eqnarray}
\Gamma\left(P_c\to \eta_c N\right)&=&31.3972 \, G_{P\eta_cN}^2 \,\,{\rm{MeV}}\, , \nonumber\\
&=&5.02\pm1.00\,{\rm{MeV}}\, , \nonumber\\
\Gamma\left(P_c\to \bar{D}^0 \Lambda_c^+\right)&=&49.4472 \, G_{P\bar{D}^0\Lambda_c^+}^2 \,\,{\rm{MeV}}\, ,\nonumber\\
&=&2.85\pm0.47\,{\rm{MeV}}\, , \nonumber\\
\Gamma\left(P_c\to J/\psi N\right)&=&29.5806 \, G_T^2 - 97.1516\,G_V G_T + 80.2825\, G_V^2\, ,\nonumber\\
&=&6.45\pm1.84\,\,{\rm{MeV}}\, .
\end{eqnarray}
If we saturate the decay width of the $P_c$ with the two-body strong decays  to the $\eta_c N$, $\bar{D}^0 \Lambda_c^+$ and $J/\psi N$, we can obtain the total width
$\Gamma(P_c)=14.32\pm3.31\,\rm{MeV}$, which is compatible with the experimental value  $\Gamma_{P_c(4312)} = 9.8\pm2.7^{+ 3.7}_{- 4.5} \mbox{ MeV}$ from the LHCb collaboration \cite{LHCb-Pc4312}.
The present calculations also support assigning the
$P_c(4312)$ to be the diquark-diquark-antiquark type hidden-charm pentaquark  state with the spin-parity $J^P={\frac{1}{2}}^-$. The $P_c(4312)$ maybe have a diquark-diquark-antiquark type pentaquark core with the typical size of the $qqq$-type baryon states, the strong couplings to the meson-baryon pairs $\bar{D}^0\Sigma_c^+$ and $\bar{D}^-\Sigma_c^{++}$  lead to some pentaquark molecule components according to the large hadronic coupling constants  $|G_{P\bar{D}^-\Sigma_c^{++}}|= \sqrt{2}|G_{P\bar{D}^0\Sigma_c^+}|\gg |G_{P\bar{D}^0\Lambda_c^+}|$, and the $P_c(4312)$ maybe spend a rather large time as the $\bar{D}^0\Sigma_c^+$ and $\bar{D}^-\Sigma_c^{++}$ molecular states, just in the case of the $f_0(980)$, $a_0(980)$ and $Y(4660)$. In Ref.\cite{Pc4312-Wang-mole-decay}, we  assign the $P_c(4312)$ to be the $\bar{D}\Sigma_c$ pentaquark molecular state with the spin-parity $J^P={\frac{1}{2}}^-$ tentatively, and  explore  its two-body strong decays with the QCD sum rules, and obtain the  partial decay  widths $\Gamma\left(P_c(4312)\to \eta_c N\right)=0.255\,{\rm{MeV}}$ and $\Gamma\left(P_c(4312)\to J/\psi N\right)=9.296\,\,{\rm{MeV}}$.
The $P_c(4312)$ has quite different  branching fractions  in the scenarios of the pentaquark  state and pentaquark molecular state.
 We can search for the $P_c(4312)$ in the $\eta_c N$, $\bar{D}^0 \Lambda_c^+$ and $J/\psi N$ invariant mass spectrum, and measure the branching fractions
${\rm Br}\left(P_c(4312)\to \eta_c N,\, \bar{D}^0 \Lambda_c^+,\, J/\psi N\right)$ precisely, which maybe shed light on the nature of the $P_c(4312)$ unambiguously,  test  the predictions of the QCD sum rules and examine the hadronic dressing mechanism. If the hadronic dressing mechanism works, the $P_c(4318)$ has both the diquark-diquark-antiquark type and meson-baryon type Fock components, we should introduce mixing effects in the interpolating current, and fix the mixing angle by the precise experimental data in the future.

\section{Conclusion}
In this article, we illustrate how to calculate the hadronic coupling constants  of the hidden-charm pentaquark states with the QCD sum rules based on rigorous  quark-hadron quality, then
study the hadronic coupling constants of the lowest diquark-diquark-antiquark type pentaquark state with the spin-parity $J^P={\frac{1}{2}}^-$  in a consistent  way. The predicted total width $\Gamma(P_c)=14.32\pm3.31 \,\rm{MeV}$ is compatible with the experimental data $\Gamma_{P_c(4312)} = 9.8\pm2.7^{+ 3.7}_{- 4.5} \mbox{ MeV}$ from the LHCb collaboration, and favors assigning the $P_c(4312)$ to be the  $[ud][uc]\bar{c}$  type compact pentaquark state with the spin-parity $J^P={\frac{1}{2}}^-$.
The $P_c(4312)$ maybe have a diquark-diquark-antiquark type pentaquark core with the typical size of the $qqq$ type baryon states, the strong couplings to the meson-baryon pairs $\bar{D}^0\Sigma_c^+$ and $\bar{D}^-\Sigma_c^{++}$  lead to some pentaquark molecule components according to the large hadronic coupling constants  $|G_{PD^-\Sigma_c^{++}}|= \sqrt{2}|G_{P\bar{D}^0\Sigma_c^+}|\gg |G_{P\bar{D}^0\Lambda_c^+}|$, just in the case of the $f_0(980)$, $a_0(980)$ and $Y(4660)$. The $P_c(4312)$ has quite different  branching fractions in the scenarios of the pentaquark  state and pentaquark molecular state,
   we can distinguish or  compromising the  two scenarios unambiguously  by  measuring the branching fractions
${\rm Br}\left(P_c(4312)\to \eta_c N,\, \bar{D}^0 \Lambda_c^+,\, J/\psi N\right)$ precisely.

\section*{Appendix}
The explicit expressions of the QCD spectral densities $\rho^{\eta_cN}_{QCD}(s,u)$, $\rho^{\bar{D}^0\Lambda_c^+}_{QCD}(s,u)$,
$\rho^{\bar{D}^0\Sigma_c^+}_{QCD}(s,u)$, $\rho^{J/\psi N,1}_{QCD}(s,u)$ and $\rho^{J/\psi N,2}_{QCD}(s,u)$,
\begin{eqnarray}\label{rho-etacN}
\rho^{\eta_cN}_{QCD}(s,u)&=&-\frac{m_{c}}{2048\pi^6} \int_{x_{i}}^{x_{f}}dx\, u^2   -\frac{m_{c}\langle\bar{q}q\rangle^2}{12\pi^2}  \int_{x_{i}}^{x_{f}}dx\, \delta\left(u\right) \nonumber\\
  &&+\frac{\langle\bar{q}g_{s}\sigma Gq\rangle}{4608\pi^4} \int_{x_{i}}^{x_{f}}dx
\,\left(1+x\right) u\,\delta\left(s-\widetilde{m}_c^2\right)+\frac{7m_c\langle \bar{q}q\rangle\langle \bar{q}g_s\sigma Gq\rangle}{192\pi^2T_{2}^{2}}\int_{x_{i}}^{x_{f}}dx\,\delta\left(u\right) \nonumber\\
&&-\frac{m_{c}\langle\bar{q}q\rangle\langle\bar{q}g_{s}\sigma Gq\rangle}{72\pi^2} \int_{x_{i}}^{x_{f}}dx\, \frac{1}{x}\,\delta\left(s-\widetilde{m}_c^2\right)\,\delta(u) \nonumber\\
 && +\frac{m_{c}^{3}}{18432\pi^4T_{1}^{4}} \langle\frac{\alpha_{s}GG}{\pi}\rangle  \int_{x_{i}}^{x_{f}}dx\, \frac{u^2}{x^3}\,\delta\left(s-\widetilde{m}_c^2\right)\nonumber\\
&&- \frac{m_{c}}{1024\pi^4} \langle\frac{\alpha_{s}GG}{\pi}\rangle\int_{x_{i}}^{x_{f}}dx- \frac{m_{c}}{4608\pi^4} \langle\frac{\alpha_{s}GG}{\pi}\rangle\int_{x_{i}}^{x_{f}}dx\, \frac{1}{x}\, u \,\delta\left(s-\widetilde{m}_c^2\right)\nonumber\\
&&+\frac{m_{c}}{6144\pi^4T_{1}^{2}} \langle\frac{\alpha_{s}GG}{\pi}\rangle  \int_{x_{i}}^{x_{f}}dx \frac{2x-1}{x^2} u^2 \delta\left(s-\widetilde{m}_c^2\right)\nonumber\\
 &&+\frac{m_{c}^{3}}{108T_{1}^{4}} \langle \bar{q}q\rangle^2\langle\frac{\alpha_{s}GG}{\pi}\rangle  \int_{x_{i}}^{x_{f}}dx\, \frac{1}{x^3}\,\delta\left(s-\widetilde{m}_c^2\right)\delta\left(u\right)\nonumber\\
&&+\frac{m_{c}}{36T_{1}^{2}} \langle \bar{q}q\rangle^2\langle\frac{\alpha_{s}GG}{\pi}\rangle  \int_{x_{i}}^{x_{f}}dx\frac{2x-1}{x^2} \delta\left(s-\widetilde{m}_c^2\right)\delta\left(u\right)\nonumber\\
 &&-\frac{m_{c}}{432T_{2}^{2}} \langle \bar{q}q\rangle^2\langle\frac{\alpha_{s}GG}{\pi}\rangle  \int_{x_{i}}^{x_{f}}dx\, \frac{1}{x}\, \delta\left(s-\widetilde{m}_c^2\right)\delta\left(u\right)\nonumber\\
&&+\frac{11m_{c}\langle\bar{q}g_{s}\sigma Gq\rangle^2}{9216\pi^2T_2^2} \int_{x_{i}}^{x_{f}}dx\, \frac{1}{x(1-x)}\, \delta\left(s-\widetilde{m}_c^2\right)\,\delta(u) \nonumber\\
&&-\frac{m_{c}\langle\bar{q}g_{s}\sigma Gq\rangle^2}{1536\pi^2T_1^2} \int_{x_{i}}^{x_{f}}dx\, \frac{1}{x(1-x)}\, \delta\left(s-\widetilde{m}_c^2\right)\,\delta(u) \, , \end{eqnarray}

\begin{eqnarray}\label{rho-D0Lambdac}
\rho^{\bar{D}^0\Lambda_c^+}_{QCD}(s,u)&=&\frac{9m_{c}}{4096\pi^6} \int_{x_{i}}^{1}dx\int_{y_{i}}^{1}dy\, (1-x)y(1-y)^2\left(u-\widetilde{m}_y^2\right)^2\nonumber\\
 &&-\frac{3\langle\bar{q}q\rangle}{1024\pi^4} \int_{y_{i}}^{1}dy\, y(1-y)^2\delta\left(s-m_c^2\right)\left(u-\widetilde{m}_y^2\right)^2\nonumber\\
&&+\frac{m_{c}^2\langle\bar{q}q\rangle}{128\pi^4} \int_{x_{i}}^{1}dx\int_{y_{i}}^{1}dy\, (1-x)(1-y)\nonumber\\
&&+\frac{\langle\bar{q}g_{s}\sigma Gq\rangle}{4096\pi^4T_1^2} \int_{y_{i}}^{1}dy\, y(1-y)^2\left(4+\frac{3s}{T_1^2}\right)\delta\left(s-m_c^2\right)\left(u-\widetilde{m}_y^2\right)^2\nonumber\\
&&+\frac{m_{c}^2\langle\bar{q}g_{s}\sigma Gq\rangle}{1024\pi^4}\int_{x_{i}}^{1}dx\, \int_{y_{i}}^{1}dy\, \frac{(1-x)(1-2y)}{y}\delta\left(u-\widetilde{m}_y^2\right)\nonumber\\
&&+\frac{3\langle\bar{q}g_{s}\sigma Gq\rangle}{8192\pi^4} \int_{y_{i}}^{1}dy\, (5y-3)(1-y)\delta\left(s-m_c^2\right)\left(u-\widetilde{m}_y^2\right)\nonumber\\
&&+\frac{\langle\bar{q}g_{s}\sigma Gq\rangle}{6144\pi^4} \int_{x_{i}}^{1}dx\,\int_{y_{i}}^{1}dy\, y(1-y)(3u-s)\delta\left(s-\widetilde{m}_x^2\right)\nonumber\\
&&-\frac{m_{c}^2\langle\bar{q}g_{s}\sigma Gq\rangle}{12288\pi^4}\int_{x_{i}}^{1}dx\, \int_{y_{i}}^{1}dy\, \frac{(2-7x)(1-y)}{x}\delta\left(s-\widetilde{m}_x^2\right)\nonumber\\
&&-\frac{m_{c}^3}{8192\pi^4T_1^4}\langle\frac{\alpha_{s}GG}{\pi}\rangle \int_{x_{i}}^{1}dx\int_{y_{i}}^{1}dy\, \frac{(1-x)y(1-y)^2}{x^3}\delta\left(s-\widetilde{m}_x^2\right)\left(u-\widetilde{m}_y^2\right)^2\nonumber\\
&&+\frac{m_{c}}{8192\pi^4T_1^2}\langle\frac{\alpha_{s}GG}{\pi}\rangle \int_{x_{i}}^{1}dx\int_{y_{i}}^{1}dy\, \frac{(3-4x)y(1-y)^2}{x^2}\delta\left(s-\widetilde{m}_x^2\right)\left(u-\widetilde{m}_y^2\right)^2\nonumber\\
&&-\frac{m_{c}^3}{4096\pi^4}\langle\frac{\alpha_{s}GG}{\pi}\rangle \int_{x_{i}}^{1}dx\int_{y_{i}}^{1}dy\,\frac{(1-x)(1-y)^2}{y^2} \delta\left(u-\widetilde{m}_y^2\right)\nonumber\\
&&+\frac{3m_{c}}{4096\pi^4}\langle\frac{\alpha_{s}GG}{\pi}\rangle \int_{x_{i}}^{1}dx\int_{y_{i}}^{1}dy\,y(1-x) \nonumber\\
&&-\frac{m_{c}}{16384\pi^4}\langle\frac{\alpha_{s}GG}{\pi}\rangle \int_{x_{i}}^{1}dx\int_{y_{i}}^{1}dy\,(1-x+y)(1-y)^2 (3u-s)\delta\left(s-\widetilde{m}_x^2\right)\nonumber\\
&&-\frac{m_{c}}{16384\pi^4}\langle\frac{\alpha_{s}GG}{\pi}\rangle \int_{x_{i}}^{1}dx\int_{y_{i}}^{1}dy\,\frac{(2x-1)(1-y)(5y-3)}{x} \delta\left(s-\widetilde{m}_x^2\right)\left(u-\widetilde{m}_y^2\right)\nonumber\\
&&+\frac{m_{c}^2\langle\bar{q}q\rangle}{3072\pi^2}\langle\frac{\alpha_{s}GG}{\pi}\rangle \int_{y_{i}}^{1}dy\,\frac{(1-y)^2}{y^2} \delta\left(s-m_c^2\right)\delta\left(u-\widetilde{m}_y^2\right)\nonumber\\
&&+\frac{m_{c}^2\langle\bar{q}q\rangle}{6144\pi^2T_1^6}\left(1-\frac{m_c^2}{2T_1^2}\right)\langle\frac{\alpha_{s}GG}{\pi}\rangle \int_{y_{i}}^{1}dy\, y(1-y)^2\delta\left(s-m_c^2\right)\left(u-\widetilde{m}_y^2\right)^2\nonumber\\
&&-\frac{m_{c}^4\langle\bar{q}q\rangle}{2304\pi^2T_1^4}\langle\frac{\alpha_{s}GG}{\pi}\rangle \int_{x_{i}}^{1}dx\int_{y_{i}}^{1}dy\, \frac{(1-x)(1-y)}{x^3}\delta\left(s-\widetilde{m}_x^2\right)\nonumber\\
&&+\frac{m_{c}^2\langle\bar{q}q\rangle}{768\pi^2T_2^2}\langle\frac{\alpha_{s}GG}{\pi}\rangle \int_{x_{i}}^{1}dx\int_{y_{i}}^{1}dy\,\frac{(1-x)(1-y)}{y^2}\left(1-\frac{u}{3T_2^2} \right) \delta\left(u-\widetilde{m}_y^2\right)\nonumber\\
&&+\frac{m_{c}^2\langle\bar{q}q\rangle}{4608\pi^2T_2^2}\langle\frac{\alpha_{s}GG}{\pi}\rangle \int_{x_{i}}^{1}dx\,(1-x)\delta\left(u-m_c^2\right)\nonumber\\
&&+\frac{m_{c}^2\langle\bar{q}q\rangle}{18432\pi^2}\langle\frac{\alpha_{s}GG}{\pi}\rangle \int_{x_{i}}^{1}dx\int_{y_{i}}^{1}dy\,\frac{ 8xy-2x-7y+4}{xy} \delta\left(s-\widetilde{m}_x^2\right)\delta\left(u-\widetilde{m}_y^2\right)\nonumber
\end{eqnarray}
\begin{eqnarray}
&&+\frac{\langle\bar{q}q\rangle}{18432\pi^2}\langle\frac{\alpha_{s}GG}{\pi}\rangle \int_{x_{i}}^{1}dx\int_{y_{i}}^{1}dy\,(5y-2)(3u-s)\delta\left(s-\widetilde{m}_x^2\right)\delta\left(u-\widetilde{m}_y^2\right)\nonumber\\
&&-\frac{\langle\bar{q}q\rangle}{1024\pi^2}\langle\frac{\alpha_{s}GG}{\pi}\rangle \int_{y_{i}}^{1}dy\,y \delta\left(s-m_c^2\right)\nonumber\\
&&+\frac{\langle\bar{q}q\rangle}{12288\pi^2T_1^2}\langle\frac{\alpha_{s}GG}{\pi}\rangle \int_{y_{i}}^{1}dy\,(5y-3)(1-y) \delta\left(s-m_c^2\right)\left(u-\widetilde{m}_y^2\right)\nonumber\\
&&+\frac{m_{c}^2\langle\bar{q}q\rangle}{768\pi^2T_1^2}\langle\frac{\alpha_{s}GG}{\pi}\rangle \int_{x_{i}}^{1}dx\int_{y_{i}}^{1}dy\,\frac{1-y}{x^2} \delta\left(s-\widetilde{m}_x^2\right)\nonumber \\
&&-\frac{m_{c}\langle\bar{q}q\rangle^2}{96\pi^2} \int_{y_{i}}^{1}dy\, (1-y)\delta\left(s-m_c^2\right)+\frac{m_{c}\langle\bar{q}q\rangle^2}{64\pi^2} \int_{x_{i}}^{1}dx\, (1-x)\delta\left(u-m_c^2\right)\nonumber\\
&&-\frac{m_{c}\langle\bar{q}q\rangle\langle\bar{q}g_{s}\sigma Gq\rangle}{3072\pi^2} \int_{y_{i}}^{1}dy\, \frac{(4-5y)}{y}\delta\left(s-m_c^2\right)\delta\left(u-\widetilde{m}_y^2\right)\nonumber\\
&&+\frac{m_{c}\langle\bar{q}q\rangle\langle\bar{q}g_{s}\sigma Gq\rangle}{9216\pi^2T_1^2} \int_{y_{i}}^{1}dy\, (1-y)\left(-5+\frac{24s}{T_1^2}\right)\delta\left(s-m_c^2\right)\nonumber\\
&&-\frac{m_{c}\langle\bar{q}q\rangle\langle\bar{q}g_{s}\sigma Gq\rangle}{128\pi^2T_2^2} \int_{x_{i}}^{1}dx\, (1-x)\left(1+\frac{u}{T_2^2}\right)\delta\left(u-m_c^2\right)\nonumber\\
&&-\frac{m_{c}\langle\bar{q}q\rangle\langle\bar{q}g_{s}\sigma Gq\rangle}{1024\pi^2} \int_{x_{i}}^{1}dx\,\delta\left(s-\widetilde{m}_x^2\right)\delta\left(u-m_c^2\right)\nonumber\\
&&-\frac{\langle\bar{q}q\rangle^3}{48} \delta\left(s-m_c^2\right)\delta\left(u-m_c^2\right)\nonumber\\
&&+\frac{m_{c}\langle\bar{q}g_{s}\sigma Gq\rangle^2}{3072\pi^2T_1^2} \int_{y_{i}}^{1}dy\,\frac{(1-2y)}{y}\left(1+\frac{s}{T_1^2}\right)\delta\left(s-m_c^2\right)\delta\left(u-\widetilde{m}_y^2\right)\nonumber\\
&&+\frac{3m_{c}\langle\bar{q}g_{s}\sigma Gq\rangle^2}{3072\pi^2T_2^8} \int_{x_{i}}^{1}dx\, (1-x)u^2\delta\left(u-m_c^2\right)\nonumber\\
&&+\frac{m_{c}\langle\bar{q}g_{s}\sigma Gq\rangle^2}{4096\pi^2T_2^2} \int_{x_{i}}^{1}dx\,\left(1+\frac{u}{T_2^2}\right)\delta\left(s-\widetilde{m}_x^2\right)\delta\left(u-m_c^2\right)\nonumber\\
&&+\frac{m_{c}\langle\bar{q}g_{s}\sigma Gq\rangle^2}{4096\pi^2T_2^2} \delta\left(s-m_c^2\right)\delta\left(u-m_c^2\right)\nonumber\\
&&-\frac{m_{c}\langle\bar{q}g_{s}\sigma Gq\rangle^2}{294912\pi^2T_1^2} \int_{y_{i}}^{1}dy\,\frac{96-191y}{y}\delta\left(s-m_c^2\right)\delta\left(u-\widetilde{m}_y^2\right)\nonumber\\
&&+\frac{5m_{c}^3\langle\bar{q}g_{s}\sigma Gq\rangle^2}{36864\pi^2T_1^6} \int_{y_{i}}^{1}dy\,(1-y) \delta\left(s-m_c^2\right)\nonumber\\
&&-\frac{m_{c}\langle\bar{q}g_{s}\sigma Gq\rangle^2}{6144\pi^2T_2^2} \int_{y_{i}}^{1}dy\,\frac{1}{y}\delta\left(s-m_c^2\right)\delta\left(u-\widetilde{m}_y^2\right)\nonumber\\
&&+\frac{7m_{c}\langle\bar{q}g_{s}\sigma Gq\rangle^2}{36864\pi^2T_1^2} \int_{x_{i}}^{1}dx\,\frac{1}{x}\delta\left(s-\widetilde{m}_x^2\right)\delta\left(u-m_c^2\right)\nonumber\\
&&+\frac{m_{c}\langle\bar{q}g_{s}\sigma Gq\rangle^2}{221184\pi^2T_2^2} \int_{x_{i}}^{1}dx\,\frac{8-37x}{x}\delta\left(s-\widetilde{m}_x^2\right)\delta\left(u-m_c^2\right)\nonumber\\
&&+\frac{m_{c}^3\langle\bar{q}q\rangle^2}{1728T_1^6}\left(1-\frac{m_c^2}{2T_1^2} \right)\langle\frac{\alpha_{s}GG}{\pi}\rangle \int_{y_{i}}^{1}dy\, (1-y)\delta\left(s-m_c^2\right)\nonumber
\end{eqnarray}
\begin{eqnarray}
&&-\frac{m_{c}^3\langle\bar{q}q\rangle^2}{1152T_1^4}\langle\frac{\alpha_{s}GG}{\pi}\rangle \int_{x_{i}}^{1}dx\,\frac{(1-x)}{x^3}\delta\left(s-\widetilde{m}_x^2\right)\delta\left(u-m_c^2\right)\nonumber\\
&&-\frac{m_{c}^3\langle\bar{q}q\rangle^2}{1152T_2^6}\left(1-\frac{m_c^2}{T_2^2} \right)\langle\frac{\alpha_{s}GG}{\pi}\rangle \int_{x_{i}}^{1}dx\,(1-x)\delta\left(u-m_c^2\right)\nonumber\\
&&-\frac{m_{c}\langle\bar{q}q\rangle^2}{576T_2^2}\langle\frac{\alpha_{s}GG}{\pi}\rangle \int_{y_{i}}^{1}dy\,\frac{(1-y)}{y^2} \left( 1-\frac{u}{3T_2^2}\right) \delta\left(s-m_c^2\right)\delta\left(u-\widetilde{m}_y^2\right)\nonumber\\
&&-\frac{m_{c}\langle\bar{q}q\rangle^2}{1536T_2^2}\langle\frac{\alpha_{s}GG}{\pi}\rangle \int_{x_{i}}^{1}dx\,\frac{1-2x}{x}
\delta\left(s-\widetilde{m}_x^2\right)\delta\left(u-m_c^2\right)\nonumber\\
&&+\frac{m_{c}\langle\bar{q}q\rangle^2}{1152T_1^2}\langle\frac{\alpha_{s}GG}{\pi}\rangle \int_{x_{i}}^{1}dx\,\frac{3-4x}{x^2}\delta\left(s-\widetilde{m}_x^2\right)\delta\left(u-m_c^2\right)\nonumber\\
&&-\frac{m_{c}\langle\bar{q}q\rangle^2}{13824T_1^2}\langle\frac{\alpha_{s}GG}{\pi}\rangle \int_{y_{i}}^{1}dy\,\frac{y+2}{y}
\delta\left(s-m_c^2\right)\delta\left(u-\widetilde{m}_y^2\right)\nonumber\\
&&-\frac{m_{c}\langle\bar{q}q\rangle^2}{3456T_2^2}\langle\frac{\alpha_{s}GG}{\pi}\rangle \delta\left(s-m_c^2\right)\delta\left(u-m_c^2\right)\, ,
\end{eqnarray}

\begin{eqnarray}\label{rho-D0Sigmac}
\rho^{\bar{D}^0\Sigma_c^+}_{QCD}(s,u)&=&-\frac{3m_{c}}{1024\pi^6} \int_{x_{i}}^{1}dx\int_{y_{i}}^{1}dy\, (1-x)y(1-y)^2\left(u-\widetilde{m}_y^2\right)^2\nonumber\\
 &&+\frac{\langle\bar{q}q\rangle}{256\pi^4} \int_{y_{i}}^{1}dy\, y(1-y)^2\delta\left(s-m_c^2\right)\left(u-\widetilde{m}_y^2\right)^2\nonumber\\
 &&+\frac{3m_{c}^2\langle\bar{q}q\rangle}{128\pi^4}\int_{x_{i}}^{1}dx \int_{y_{i}}^{1}dy\, (1-x)(1-y)\nonumber\\
&&+\frac{\langle\bar{q}g_{s}\sigma Gq\rangle}{512\pi^4} \int_{y_{i}}^{1}dy\, y(1-y)\delta\left(s-m_c^2\right)\left(u-\widetilde{m}_y^2\right)\nonumber\\
&&+\frac{m_{c}^2\langle\bar{q}g_{s}\sigma Gq\rangle}{1024\pi^4} \int_{x_{i}}^{1}dx\,\int_{y_{i}}^{1}dy\,(1-y)\delta\left(s-\widetilde{m}_x^2\right)\nonumber\\
&&+\frac{m_{c}^2\langle\bar{q}g_{s}\sigma Gq\rangle}{1024\pi^4} \int_{x_{i}}^{1}dx\,\int_{y_{i}}^{1}dy\,\frac{(1-x)(6-13y)}{y} \delta\left(u-\widetilde{m}_y^2\right)\nonumber\\
&&-\frac{m_c^2\langle\bar{q}g_{s}\sigma Gq\rangle}{1024\pi^4T_1^4} \int_{y_{i}}^{1}dy\, y(1-y)^2\,\delta\left(s-m_c^2\right)\left(u-\widetilde{m}_y^2\right)^2\nonumber\\
&&-\frac{m_{c}}{2048\pi^4T_1^2}\left(1-\frac{m_c^2}{3T_1^2} \right)\langle\frac{\alpha_{s}GG}{\pi}\rangle \int_{x_{i}}^{1}dx\,\int_{y_{i}}^{1}dy\,\frac{(1-x)y(1-y)^2}{x^2} \delta\left(s-\widetilde{m}_x^2\right)\left(u-\widetilde{m}_y^2\right)^2\nonumber\\
&&+\frac{m_{c}^3}{3072\pi^4}\langle\frac{\alpha_{s}GG}{\pi}\rangle \int_{x_{i}}^{1}dx\,\int_{y_{i}}^{1}dy\,\frac{(1-x)(1-y)^2}{y^2}\delta\left(u-\widetilde{m}_y^2\right)\nonumber\\
&&-\frac{m_{c}}{1024\pi^4}\langle\frac{\alpha_{s}GG}{\pi}\rangle \int_{x_{i}}^{1}dx\,\int_{y_{i}}^{1}dy\,(1-x)(1-y) \nonumber\\
&&-\frac{m_{c}}{2048\pi^4T_1^2}\langle\frac{\alpha_{s}GG}{\pi}\rangle \int_{x_{i}}^{1}dx\,\int_{y_{i}}^{1}dy\,\frac{y(1-y)^2}{x}\delta\left(s-\widetilde{m}_x^2\right)\left(u-\widetilde{m}_y^2\right)^2\nonumber\\
&&+\frac{m_{c}}{3072\pi^4}\langle\frac{\alpha_{s}GG}{\pi}\rangle  \int_{x_{i}}^{1}dx\,\int_{y_{i}}^{1}dy\,\frac{(1-2x)y(1-y)}{x} \delta\left(s-\widetilde{m}_x^2\right)\left(u-\widetilde{m}_y^2\right)\nonumber\\
&&-\frac{m_{c}}{73728\pi^4}\langle\frac{\alpha_{s}GG}{\pi}\rangle  \int_{x_{i}}^{1}dx\,\int_{y_{i}}^{1}dy\,\frac{(y+2)(1-y)^2}{y}(3u-s) \delta\left(s-\widetilde{m}_x^2\right)\nonumber
\end{eqnarray}
\begin{eqnarray}
&&-\frac{m_{c}^2\langle\bar{q}q\rangle}{2304\pi^2}\langle\frac{\alpha_{s}GG}{\pi}\rangle \int_{y_{i}}^{1}dy\,\frac{(1-y)^2}{y^2} \delta\left(s-m_c^2\right)\delta\left(u-\widetilde{m}_y^2\right)\nonumber\\
&&+\frac{\langle\bar{q}q\rangle}{768\pi^2}\langle\frac{\alpha_{s}GG}{\pi}\rangle \int_{y_{i}}^{1}dy\,(1-y) \delta\left(s-m_c^2\right)\nonumber\\
&&+\frac{m_{c}^2\langle\bar{q}q\rangle}{1536\pi^2T_2^2}\langle\frac{\alpha_{s}GG}{\pi}\rangle\int_{x_{i}}^{1}dx\, \int_{y_{i}}^{1}dy\,\frac{(1-x)}{y} \delta\left(u-\widetilde{m}_y^2\right)\nonumber\\
&&+\frac{\langle\bar{q}q\rangle}{2304\pi^2T_1^2}\langle\frac{\alpha_{s}GG}{\pi}\rangle \int_{y_{i}}^{1}dy\,y(1-y) \delta\left(s-m_c^2\right)\left(u-\widetilde{m}_y^2\right)\nonumber\\
&&+\frac{m_{c}^2\langle\bar{q}q\rangle}{256\pi^2T_1^2}\langle\frac{\alpha_{s}GG}{\pi}\rangle  \int_{x_{i}}^{1}dx\,\int_{y_{i}}^{1}dy\,\frac{(1-y)}{x} \delta\left(s-\widetilde{m}_x^2\right)\nonumber\\
&&-\frac{m_{c}^2\langle\bar{q}q\rangle}{4608\pi^2T_1^6}\left(1-\frac{m_c^2}{2T_1^2} \right)\langle\frac{\alpha_{s}GG}{\pi}\rangle \int_{y_{i}}^{1}dy\,y(1-y)^2 \delta\left(s-m_c^2\right)\left(u-\widetilde{m}_y^2\right)^2\nonumber\\
&&+\frac{m_{c}^2\langle\bar{q}q\rangle}{1536\pi^2T_2^2}\langle\frac{\alpha_{s}GG}{\pi}\rangle \int_{x_{i}}^{1}dx\,(1-x)\delta\left(u-m_c^2\right)\nonumber\\
&&+\frac{\langle\bar{q}q\rangle}{9216\pi^2}\langle\frac{\alpha_{s}GG}{\pi}\rangle \int_{x_{i}}^{1}dx\,\int_{y_{i}}^{1}dy\,(1+y)(3u-s) \delta\left(s-\widetilde{m}_x^2\right)\delta\left(u-\widetilde{m}_y^2\right)\nonumber\\
&&+\frac{m_{c}^2\langle\bar{q}q\rangle}{3072\pi^2}\langle\frac{\alpha_{s}GG}{\pi}\rangle \int_{x_{i}}^{1}dx\,\int_{y_{i}}^{1}dy\,\frac{x}{y}\, \delta\left(s-\widetilde{m}_x^2\right)\delta\left(u-\widetilde{m}_y^2\right)\nonumber\\
&&+\frac{m_{c}^2\langle\bar{q}q\rangle}{256\pi^2T_1^2}\langle\frac{\alpha_{s}GG}{\pi}\rangle \int_{x_{i}}^{1}dx\,\int_{y_{i}}^{1}dy\,\frac{(1-x)(1-y)}{x^2} \left( 1-\frac{s}{3T_1^2}\right) \delta\left(s-\widetilde{m}_x^2\right)\nonumber\\
&&+\frac{m_{c}^2\langle\bar{q}q\rangle}{256\pi^2T_2^2}\langle\frac{\alpha_{s}GG}{\pi}\rangle \int_{x_{i}}^{1}dx\,\int_{y_{i}}^{1}dy\,\frac{(1-x)(1-y)}{y^2} \left(1-\frac{u}{3T_2^2}\right) \delta\left(u-\widetilde{m}_y^2\right)\nonumber\\
&&-\frac{m_{c}\langle\bar{q}q\rangle^2}{32\pi^2} \int_{y_{i}}^{1}dy\, (1-y)\delta\left(s-m_c^2\right)-\frac{m_{c}\langle\bar{q}q\rangle^2}{96\pi^2} \int_{x_{i}}^{1}dx\, (1-x)\delta\left(u-m_c^2\right)\nonumber\\
&&+\frac{m_{c}\langle\bar{q}q\rangle\langle\bar{q}g_{s}\sigma Gq\rangle}{384\pi^2T_2^2} \int_{x_{i}}^{1}dx\, (1-x)\left(1+\frac{2u}{T_2^2}\right)\delta\left(u-m_c^2\right)\nonumber\\
&&+\frac{m_{c}\langle\bar{q}q\rangle\langle\bar{q}g_{s}\sigma Gq\rangle}{1536\pi^2} \int_{y_{i}}^{1}dy\,\frac{25y-14}{y} \delta\left(s-m_c^2\right)\delta\left(u-\widetilde{m}_y^2\right)\nonumber\\
&&-\frac{m_{c}\langle\bar{q}q\rangle\langle\bar{q}g_{s}\sigma Gq\rangle}{768\pi^2T_1^2} \int_{y_{i}}^{1}dy\, (1-y)\left(1-\frac{6s}{T_1^2}\right)\delta\left(s-m_c^2\right)\nonumber\\
&&-\frac{m_{c}\langle\bar{q}q\rangle\langle\bar{q}g_{s}\sigma Gq\rangle}{4608\pi^2} \int_{x_{i}}^{1}dx\,\frac{9-2x}{x}\delta\left(s-\widetilde{m}_x^2\right) \delta\left(u-m_c^2\right)\nonumber\\
&&-\frac{m_{c}\langle\bar{q}q\rangle^2}{576T_1^2}\langle\frac{\alpha_{s}GG}{\pi}\rangle \int_{x_{i}}^{1}dx\,\frac{1}{x} \delta\left(s-\widetilde{m}_x^2\right)\delta\left(u-m_c^2\right)\nonumber\\
&&-\frac{m_{c}\langle\bar{q}q\rangle^2}{1152T_2^2}\langle\frac{\alpha_{s}GG}{\pi}\rangle \int_{y_{i}}^{1}dy\,\frac{1}{y} \delta\left(s-m_c^2\right)\delta\left(u-\widetilde{m}_y^2\right)\nonumber\\
&&-\frac{m_{c}\langle\bar{q}q\rangle^2}{1152T_2^2}\langle\frac{\alpha_{s}GG}{\pi}\rangle \delta\left(s-m_c^2\right)\delta\left(u-m_c^2\right)\nonumber\\
&&-\frac{m_{c}\langle\bar{q}q\rangle^2}{576T_1^2}\langle\frac{\alpha_{s}GG}{\pi}\rangle \int_{x_{i}}^{1}dx\,\frac{1-x}{x^2} \left( 1-\frac{s}{3T_1^2}\right) \delta\left(s-\widetilde{m}_x^2\right)\delta\left(u-m_c^2\right)\nonumber
\end{eqnarray}
\begin{eqnarray}
&&-\frac{m_{c}\langle\bar{q}q\rangle^2}{192T_2^2}\langle\frac{\alpha_{s}GG}{\pi}\rangle \int_{y_{i}}^{1}dy\,\frac{1-y}{y^2}\left(1-\frac{u}{3T_2^2} \right) \delta\left(s-m_c^2\right) \delta\left(u-\widetilde{m}_y^2\right)\nonumber\\
&&+\frac{m_{c}\langle\bar{q}q\rangle^2}{6912T_2^2}\langle\frac{\alpha_{s}GG}{\pi}\rangle \int_{x_{i}}^{1}dx\,\frac{1-2x}{x} \delta\left(s-\widetilde{m}_x^2\right)\delta\left(u-m_c^2\right)\nonumber\\
&&-\frac{m_{c}\langle\bar{q}q\rangle^2}{2304T_1^2}\langle\frac{\alpha_{s}GG}{\pi}\rangle \int_{y_{i}}^{1}dy\,\frac{1-2y}{y}
\delta\left(s-m_c^2\right)\delta\left(u-\widetilde{m}_y^2\right)\nonumber\\
&&+\frac{m_{c}^3\langle\bar{q}q\rangle^2}{576T_1^6}\left(1-\frac{m_c^2}{2T_1^2}  \right)\langle\frac{\alpha_{s}GG}{\pi}\rangle \int_{y_{i}}^{1}dy\,(1-y) \delta\left(s-m_c^2\right)\nonumber\\
&&+\frac{m_{c}^3\langle\bar{q}q\rangle^2}{1728T_2^6}\left( 1-\frac{m_c^2}{T_2^2}\right)\langle\frac{\alpha_{s}GG}{\pi}\rangle \int_{x_{i}}^{1}dx\,(1-x) \delta\left(u-m_c^2\right)\nonumber\\
&&+\frac{\langle\bar{q}q\rangle^3}{72} \delta\left(s-m_c^2\right)\delta\left(u-m_c^2\right)+\frac{m_{c}\langle\bar{q}g_{s}\sigma Gq\rangle^2}{2048T_2^2}  \delta\left(s-m_c^2\right) \delta\left(u-m_c^2\right)\nonumber\\
&&+\frac{m_{c}\langle\bar{q}g_{s}\sigma Gq\rangle^2}{512\pi^2T_1^2} \int_{y_{i}}^{1}dy\, \delta\left(s-m_c^2\right)+\frac{m_{c}^3\langle\bar{q}g_{s}\sigma Gq\rangle^2}{3072\pi^2T_1^6} \int_{y_{i}}^{1}dy\,(1-y) \delta\left(s-m_c^2\right)\nonumber\\
&&-\frac{m_{c}\langle\bar{q}g_{s}\sigma Gq\rangle^2}{18432\pi^2T_1^2} \int_{x_{i}}^{1}dx\,\frac{1}{x}\delta\left(s-\widetilde{m}_x^2\right)\delta\left(u-m_c^2\right)\nonumber\\
&&+\frac{m_{c}\langle\bar{q}g_{s}\sigma Gq\rangle^2}{36864\pi^2T_2^2} \int_{x_{i}}^{1}dx\,\frac{9-2x}{x}\left(1+\frac{2u}{T_2^2}\right)\delta\left(s-\widetilde{m}_x^2\right) \delta\left(u-m_c^2\right)\nonumber\\
&&+\frac{m_{c}\langle\bar{q}g_{s}\sigma Gq\rangle^2}{3072\pi^2T_1^2} \int_{y_{i}}^{1}dy\,\frac{6-13y}{y}\left(1+\frac{s}{T_1^2}\right) \delta\left(s-m_c^2\right)\delta\left(u-\widetilde{m}_y^2\right)\nonumber\\
&&-\frac{m_{c}\langle\bar{q}g_{s}\sigma Gq\rangle^2}{9216\pi^2T_2^2} \int_{y_{i}}^{1}dy\,\frac{1}{y}\delta\left(s-m_c^2\right)\delta\left(u-\widetilde{m}_y^2\right)\nonumber\\
&&+\frac{m_{c}\langle\bar{q}g_{s}\sigma Gq\rangle^2}{147456\pi^2T_1^2} \int_{y_{i}}^{1}dy\,\frac{379y-326}{y}\delta\left(s-m_c^2\right)\delta\left(u-\widetilde{m}_y^2\right)\nonumber\\
&&+\frac{m_{c}^3\langle\bar{q}g_{s}\sigma Gq\rangle^2}{1536\pi^2T_2^6}\left(1-\frac{m_c^2}{T_2^2} \right) \int_{x_{i}}^{1}dx\,(1-x)\delta\left(u-m_c^2\right)\, ,
\end{eqnarray}

\begin{eqnarray}\label{rho-JpsiN1}
\rho^{J/\psi N,1}_{QCD}(s,u)&=&-\frac{m_{c}}{2048\pi^6} \int_{x_{i}}^{x_{f}}dx\, u^2   -\frac{m_{c}\langle\bar{q}q\rangle^2}{12\pi^2}  \int_{x_{i}}^{x_{f}}dx\, \delta\left(u\right)\nonumber\\
&&+\frac{\langle\bar{q}g_{s}\sigma Gq\rangle}{9216\pi^4}\int_{x_{i}}^{x_{f}}dx\,\left(1+x\right)u\delta\left(s-\widetilde{m}_c^2\right)+\frac{7m_{c}\langle\bar{q}q\rangle\langle\bar{q}g_{s}\sigma Gq\rangle}{192\pi^2T_2^2} \int_{x_{i}}^{x_{f}}dx\,\delta\left(u\right)\nonumber\\
&&+\frac{m_{c}\langle\bar{q}q\rangle\langle\bar{q}g_{s}\sigma Gq\rangle}{576\pi^2} \int_{x_{i}}^{x_{f}}dx\,\frac{3x-5}{x}\delta\left(s-\widetilde{m}_c^2\right)\delta\left(u\right)-\frac{m_{c}}{1024\pi^4}\langle\frac{\alpha_{s}GG}{\pi}\rangle \int_{x_{i}}^{x_{f}}dx\,\nonumber\\
&&-\frac{m_{c}}{9216\pi^4}\langle\frac{\alpha_{s}GG}{\pi}\rangle \int_{x_{i}}^{x_{f}}dx\,\frac{1}{x(1-x)}u\left(1-\frac{u}{T_1^2}\right)\delta\left(s-\widetilde{m}_c^2\right)\nonumber\\
&&-\frac{m_{c}}{6144\pi^4T_1^2}\langle\frac{\alpha_{s}GG}{\pi}\rangle \int_{x_{i}}^{x_{f}}dx\,\frac{1}{x^2}\left( 1-\frac{m_c^2}{3xT_1^2} \right)u^2\delta\left(s-\widetilde{m}_c^2\right)\nonumber\\
&&+\frac{m_{c}^3\langle\bar{q}q\rangle^2}{108T_1^4}\langle\frac{\alpha_{s}GG}{\pi}\rangle \int_{x_{i}}^{x_{f}}dx\,\frac{1}{x^3}\delta\left(s-\widetilde{m}_c^2\right)\delta\left(u\right)\nonumber\\
&&-\frac{m_{c}\langle\bar{q}q\rangle^2}{432T_2^2}\langle\frac{\alpha_{s}GG}{\pi}\rangle \int_{x_{i}}^{x_{f}}dx\,\frac{1}{x}\delta\left(s-\widetilde{m}_c^2\right)\delta(u)\nonumber
\end{eqnarray}
\begin{eqnarray}
&&+\frac{m_{c}\langle\bar{q}q\rangle^2}{216T_1^2}\langle\frac{\alpha_{s}GG}{\pi}\rangle \int_{x_{i}}^{x_{f}}dx\,\frac{6x^2-13x-6}{x^2(x-1)}\delta\left(s-\widetilde{m}_c^2\right)\delta\left(u\right)\nonumber\\
&&+\frac{m_{c}\langle\bar{q}g_{s}\sigma Gq\rangle^2}{4608\pi^2T_2^2} \int_{x_{i}}^{x_{f}}dx\,\frac{7-4x}{x}\delta\left(s-\widetilde{m}_c^2\right)\delta\left(u\right)\nonumber\\
&&+\frac{m_{c}\langle\bar{q}g_{s}\sigma Gq\rangle^2}{221184\pi^2T_1^2} \int_{x_{i}}^{x_{f}}dx\,\frac{61x+40}{x(1-x)}\delta\left(s-\widetilde{m}_c^2\right)\delta\left(u\right)\, ,
\end{eqnarray}

\begin{eqnarray}\label{rho-JpsiN2}
\rho^{J/\psi N,2}_{QCD}(s,u)&=&\frac{1}{2048\pi^6} \int_{x_{i}}^{x_{f}}dx\, \left[xs+x(1-x)\left(s-\widetilde{m}_c^2\right)   \right]u^2\nonumber\\
&&+\frac{\langle\bar{q}q\rangle^2}{12\pi^2} \int_{x_{i}}^{x_{f}}dx\,\left[xs+x(1-x)\left(s-\widetilde{m}_c^2\right) \right]\delta\left(u\right)\nonumber\\
&&-\frac{7\langle\bar{q}q\rangle\langle\bar{q}g_{s}\sigma Gq\rangle}{192\pi^2T_2^2} \int_{x_{i}}^{x_{f}}dx\,\left[xs+x(1-x)\left(s-\widetilde{m}_c^2\right) \right]\delta\left(u\right)\nonumber\\
&&+\frac{\langle\bar{q}q\rangle\langle\bar{q}g_{s}\sigma Gq\rangle}{144\pi^2} \int_{x_{i}}^{x_{f}}dx\,\left[s\delta\left(s-\widetilde{m}_c^2\right)+1 \right]\delta\left(u\right)\nonumber\\
&&+\frac{m_{c}^2}{18432\pi^4T_1^2}\langle\frac{\alpha_{s}GG}{\pi}\rangle \int_{x_{i}}^{x_{f}}dx\,\left[\frac{1}{x^2}\left(1-\frac{s}{T_1^2}\right)+\frac{1}{x} \right]u^2\delta\left(s-\widetilde{m}_c^2\right)\nonumber\\
&&+\frac{1}{1024\pi^4}\langle\frac{\alpha_{s}GG}{\pi}\rangle \int_{x_{i}}^{x_{f}}dx\,\left[xs+x(1-x)\left(s-\widetilde{m}_c^2\right) \right]\nonumber\\
&&+\frac{1}{2304\pi^4}\langle\frac{\alpha_{s}GG}{\pi}\rangle \int_{x_{i}}^{x_{f}}dx\,\left[s\delta\left(s-\widetilde{m}_c^2\right)+1\right]u\nonumber\\
&&+\frac{1}{36864\pi^4}\langle\frac{\alpha_{s}GG}{\pi}\rangle \int_{x_{i}}^{x_{f}}dx\,\left[1+\frac{s}{T_1^2}-\frac{s}{2x(1-x)T_1^2} \right]u^2\delta\left(s-\widetilde{m}_c^2\right)\nonumber\\
&&+\frac{m_{c}^2\langle\bar{q}q\rangle^2}{108T_1^2}\langle\frac{\alpha_{s}GG}{\pi}\rangle \int_{x_{i}}^{x_{f}}dx\,\left[\frac{1}{x^2}\left(1-\frac{s}{T_1^2}\right)+\frac{1}{x} \right]\delta\left(s-\widetilde{m}_c^2\right)\delta\left(u\right)\nonumber\\
&&+\frac{\langle\bar{q}q\rangle^2}{216T_2^2}\langle\frac{\alpha_{s}GG}{\pi}\rangle \int_{x_{i}}^{x_{f}}dx\,s\delta\left(s-\widetilde{m}_c^2\right)\delta\left(u\right)+\frac{\langle\bar{q}q\rangle^2}{216T_2^2}\langle\frac{\alpha_{s}GG}{\pi}\rangle \int_{x_{i}}^{x_{f}}dx\,\delta\left(u\right)\nonumber\\
&&+\frac{\langle\bar{q}q\rangle^2}{216}\langle\frac{\alpha_{s}GG}{\pi}\rangle \int_{x_{i}}^{x_{f}}dx\,\left[1+\frac{s}{T_1^2}-\frac{s}{2x(1-x)T_1^2} \right]\delta\left(s-\widetilde{m}_c^2\right)\delta\left(u\right)\nonumber\\
&&-\frac{3\langle\bar{q}g_{s}\sigma Gq\rangle^2}{2304\pi^2T_2^2} \int_{x_{i}}^{x_{f}}dx\,\left[s\delta\left(s-\widetilde{m}_c^2\right)+1 \right]\delta\left(u\right)\nonumber\\
&&+\frac{\langle\bar{q}g_{s}\sigma Gq\rangle^2}{110592\pi^2} \int_{x_{i}}^{x_{f}}dx\,\left[7-\frac{44s}{T_1^2}-\frac{14s}{x(1-x)T_1^2} \right]\delta\left(s-\widetilde{m}_c^2\right)\delta\left(u\right)\, ,
\end{eqnarray}
where  $x_{f}=\frac{1+\sqrt{1-4m_{c}^{2}/s}}{2}$, $x_{i}=\frac{1-\sqrt{1-4m_{c}^{2}/s}}{2}$,   $\widetilde{m}_{c}^{2}=\frac{m_{c}^{2}}{x(1-x)}$ in Eq.\eqref{rho-etacN} and Eqs.\eqref{rho-JpsiN1}-\eqref{rho-JpsiN2},
$x_i=\frac{m_c^2}{s}$, $y_i=\frac{m_c^2}{u}$, $\widetilde{m}_x^2=\frac{m_c^2}{x}$, $\widetilde{m}_y^2=\frac{m_c^2}{y}$ in Eqs.\eqref{rho-D0Lambdac}-\eqref{rho-D0Sigmac}, $\int_{x_{i}}^{x_{f}}dx\rightarrow\int_{0}^{1} dx$, $\int_{x_{i}}^{1}dx\rightarrow\int_{0}^{1} dx$ and $\int_{y_{i}}^{1}dy\rightarrow\int_{0}^{1} dy$, when the $\delta$ functions $\delta(s-\widetilde{m}_{c}^{2})$, $\delta(s-\widetilde{m}_{x}^{2})$ and $\delta(u-\widetilde{m}_{y}^{2})$ appear.

\section*{Acknowledgements}
This  work is supported by National Natural Science Foundation, Grant Number  11775079.

\end{document}